# Non-stoichiometry in SnS: How it affects thin-film morphology and electrical properties

Taichi Nogami[1], Issei Suzuki*[,1], Daiki Motai[1],
Hiroshi Tanimura[2], Tetsu Ichitsubo[2], Takahisa Omata[1]

1. Institute of Multidisciplinary Research for Advanced Materials, Tohoku University, Sendai 980-8577, Japan

2. Institute for Materials Research, Tohoku University, Sendai 980-8577, Japan

*Corresponding author: issei.suzuki@tohoku.ac.jp

## Abstract

Tin sulfide (SnS) has garnered much attention as a promising material for various applications, including solar cells and thermoelectric devices, owing to its favorable optical and electronic properties and the abundant and nontoxic nature of its constituent elements. Herein, we investigated the effect of non-stoichiometry on the morphology and electrical properties of SnS thin films. Using a unique sputtering technique with a sulfur plasma supply, SnS films with precise sulfur content control, [S]/([Sn] + [S]) ($x_S$) ranging from 0.47 to 0.51, were fabricated. Systematic characterization revealed that non-stoichiometry on the S-rich side led to a marked increase in the carrier density of p-type conduction, which was attributed to the formation of intrinsic acceptor-type defects. In contrast, non-stoichiometry on the S-poor side hardly affects the p-type electrical properties, apparently because of the self-compensation between the intrinsic acceptor- and donor-type defects. In addition, non-stoichiometry has been identified as the cause of thin-film morphological changes, with non-stoichiometric films exhibiting rough and porous surfaces. Achieving a stoichiometric composition results in smooth and dense thin-film morphologies, which are crucial for optimizing SnS thin films for device applications. These findings underscore the importance of compositional control for tailoring the morphology and electrical behavior of SnS, paving the way for more efficient SnS-based devices.

## 1. Introduction

Tin sulfide (SnS) is an emerging material that has garnered increasing attention for a variety of applications, including solar cells[1,2], thermoelectric conversion devices[3,4], light sensors[5], and transistors[6], owing to its superior optical and electronic properties and the abundance and non-toxicity of its constituent elements. The physical properties of semiconductors are influenced by intrinsic defects such as vacancies and antisites, which eventually determine the device performance. Consequently, understanding the effects of intrinsic defects on the physical properties of SnS and developing thin-film fabrication techniques that can control these defects are crucial.

Studies on the intrinsic defects in SnS have conventionally been conducted using first-principles calculations, which can quantitatively evaluate the defect formation enthalpy, carrier density, and electronic states.[7–10] These calculations are based on thermodynamically equilibrium conditions, assuming that the defect states and compositions are thermodynamically stable. However, in practical thin-film deposition processes such as sputtering and pulsed laser deposition (PLD), defect states and compositions can differ from the equilibrium conditions because particles under non-equilibrium states are rapidly quenched on the substrate[11]. Therefore, the effects of intrinsic defects in practical materials do not necessarily align with the results of first-principles calculations, making experimental evaluations indispensable.

In this study, we precisely controlled the sulfur content, which has traditionally been challenging because of its high vapor pressure, using an original sputtering process that supplied sulfur plasma (S-plasma) during deposition. This process enables the fabrication of SnS thin films with various compositions, [S]/([Sn] + [S]) ($x_S$) of $0.47 \leq x_S \leq 0.51$. We systematically investigated the effect of compositional deviations in SnS thin films on their properties such as crystal phase, preferential orientation, morphology, and electrical properties. Notably, S-poor deviations, which tend to occur because of the high sulfur vapor pressure during the SnS thin-film deposition process, result in numerous intrinsic defects that cause Fermi-level pinning and carrier scattering. We clarified that acceptor- and donor-type defects resulting from sulfur deficiencies do not significantly affect the carrier density because they mutually compensate. Additionally, while the morphology of thin films is generally influenced by fabrication parameters such as substrate temperature and deposition pressure, we found that in SnS thin films, slight deviations from stoichiometry are the primary cause of morphological changes. These findings emphasize the importance of achieving the stoichiometric composition of SnS thin films and offer valuable insights for the future development of SnS-based thin-film devices.

## 2. Experimental

### 2.1 Sample preparation

The fabrication system used in this study comprised a sputtering cathode with an SnS-sintered target and an S-plasma source for generating sulfur in the plasma state, as shown in Figure 1. A 1-inch undoped SnS sinter (99.99%, Advanced Engineering Materials Limited, China) and 40 mm square $SiO_2$ glass were used as the sputtering target and substrate, respectively. The deposition chamber was evacuated to 5 × $10^{-5}$ Pa before deposition, and the substrate was heated to 300 °C during the deposition process. A pressure gradient sputtering cathode (PGS cathode, Kenix Co. Ltd., Japan)[12], whereAr gas is supplied to the vicinity of the target surface to maintain the plasma, was used for sputtering. The Ar gas flow rate was 10 sccm, and an RF (radio frequency) power of 10 W was applied. The deposition was conducted for 3 h. The S-plasma was generated using a plasma-cracking cell (Kenix Co., Ltd.). This equipment first generates sulfur vapor by heating sulfur powder to the sulfur source temperature ($T_{\mathrm{Sulfur}}$) and then generates S-plasma by applying RF to a mixture of sulfur vapor and Ar. The generated S-plasma was conveyed to the thin-film deposition area during SnS sputtering by supplied Ar flow. Details of the sulfur evaporation rate and plasma emissions of the S-plasma under these operating conditions are provided in Figure S1 of the Supplementary Material.

### 2.2 Characterization

The crystal phases of the fabricated thin films were investigated using X-ray diffraction (XRD; SmartLab, Rigaku, Japan) and Raman spectroscopy (inVia, Renishaw plc., UK). The composition of the SnS thin films was determined using an electron probe microanalyzer (EPMA; JCA-8530F, JEOL Ltd., Japan). The surface and cross-sectional morphologies of the SnS thin films were observed by scanning electron microscopy (SEM; JCA-8630F and JSM 7800F, JEOL Ltd.). The electrical properties, including the carrier type (p- or n-type), carrier density, and mobility, were analyzed using a Hall effect measurement system (ResiTest 8300, TOYO Corporation, Japan) with Au electrodes. Au electrodes (approximately 100 nm thick) were deposited by desktop magnetron sputtering (JFC-1600, JEOL Ltd.) at 20 mA. To minimize the heating effect from sputtering and the associated chemical reactions, 32 cycles were repeated, each consisting of 10 s of Au sputtering followed by a 20 s pause for cooling. The temperature dependence of the electrical conductivity was studied using a closed-cycle He cryostat (TY-01, IWATANI Corp., Japan) using the standard 4 probe method. The optical properties of the thin films were measured using spectroscopic ellipsometry (SE-2000, Semilab Inc., Hungary), and the results were analyzed using optical modeling software (Spectroscopic Ellipsometry Analyzer v1.7, Semilab Inc.).

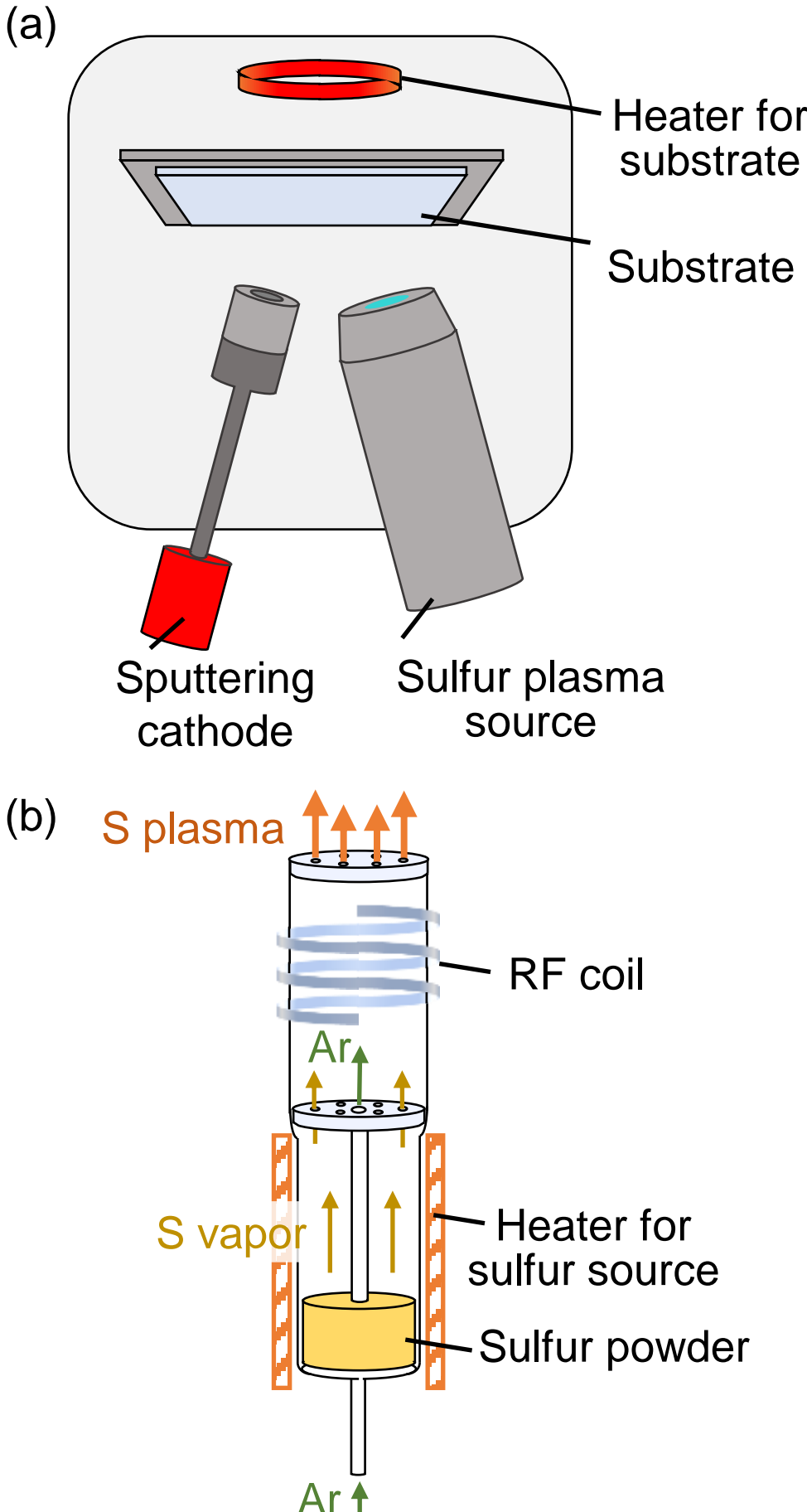

Figure 1. (a) Schematics of the deposition system composed of a sputtering cathode and S-plasma source. (b) Internal structure of the S-plasma source.

## 3. Results and discussion

### 3.1 Compositions and phases of the SnS thin films

Figure 2 shows the compositions determined by EPMA and Raman spectroscopy for thin films fabricated without a S-plasma supply, as well as those fabricated with an S-plasma supply at various $T_{Sulfur}$. The $x_S$ value of the SnS thin film fabricated without the S-plasma supply was 0.47, indicating a sulfur-poor composition. Supplying S-plasma increased the sulfur content in the thin films, with a higher $T_{Sulfur}$ resulting in greater sulfur content. This is because the high $T_{Sulfur}$ increases the amount of sulfur vapor, which in turn increases the amount of S-plasma supplied to the thin-film deposition area. The composition was closest to stoichiometric at $T_{Sulfur}$ = 100 °C, while it became S-rich at $T_{Sulfur}$ = 105 °C.

The Raman spectra of the thin films fabricated under all the conditions show peaks consistent with those reported for SnS[13]. Figure 3 shows the compositional dependence of the full width at half maximum (FWHMs) and peak positions for four representative peaks obtained from spectral fitting (see the detailed fitting procedure in Figure S2 of the Supporting Information). The thin films fabricated without the S-plasma supply exhibited the narrowest FHWM, whereas those fabricated with the S-plasma supply tended to show broader peaks. The FWHM became narrower as the composition of the thin films fabricated with the S-plasma supply approached stoichiometry. The peaks at 95 and 161 cm$^{-1}$ showed no compositional dependence, whereas the peaks at 190 and 217 cm$^{-1}$ exhibited significant red and blue shifts, respectively, for the thin films with $x_S$ = 0.49 ($T_{Sulfur}$ = 90 °C) and 0.51. These changes in the FWHM and peak positions are discussed later in comparison with the changes in the lattice parameters.

Thin films with $x_S$ = 0.47, 0.49, and 0.51 exhibited a minor peak at approximately 310 cm$^{-1}$. This peak is consistent with the main peak of $Sn_2S_3$ or $SnS_2$[14], suggesting that these thin films contain phases that include tetravalent Sn. In contrast, the spectrum of the thin film with $x_S$ = 0.5 did not exhibit the peak associated with $Sn_2S_3$ or $SnS_2$, indicating this film is a single-phase SnS. Although $Sn_2S_3$ and $SnS_2$ are sulfides with a higher ratio than SnS, the reason for their formation in thin films with sulfur-poor compositions remains unclear. It has been reported that the formation energies of $Sn_2S_3$ and $SnS_2$ are lower than that of SnS at 1100 K[15] and that $Sn_2S_3$ and $SnS_2$ readily form on the surfaces of the SnS grains[16]. The thin films with $x_S$ = 0.49 and 0.51 exhibit a morphology with sparsely deposited grains, as will be described in Section 3.2, which may have facilitated the formation of secondary phases on the grain surfaces. Because $Sn_2S_3$ and $SnS_2$ are intrinsically n-type semiconductors, it has been suggested that their inclusion as secondary phases within intrinsically p-type SnS can significantly impede proper carrier transport.[15] Therefore, from the perspective of phase purity, supplying S-plasma to achieve single-phase SnS is beneficial when using SnS as a functional material in applications such as photovoltaic and thermoelectric devices.

Figure 4(a) shows the out-of-plane XRD profiles of the fabricated thin films. All XRD peaks of the fabricated thin films were indexed to SnS, and no secondary phases were observed. The secondary phases of $Sn_2S_3$ or $SnS_2$ observed in the Raman spectra of the non-stoichiometric thin films (Figure 2) are assumed to be present in trace amounts below the detection limit of XRD. In the SnS thin film deposited without S-plasma, while the 210 peak exhibited a comparably high intensity, nearly all diffraction peaks were observed, indicating that the film was almost non-oriented SnS. In contrast, 200, 400, and 800 diffraction peaks were observed for the thin films fabricated with S-plasma supply. Additionally, 111 and 222 diffraction peaks were observed in the thin films with $x_S$ = 0.49 ($T_{Sulfur}$ = 95 °C) and 0.51 (see the detailed XRD profiles and peak analysis in Figure S3 of the Supplementary Material). In the in-plane XRD profiles shown in Figure 4(b), the thin films with non-stoichiometric compositions exhibited an almost powder pattern, while the thin film with a stoichiometric compositionexhibited strong peaks corresponding to the (0*kℓ*) planes, which are perpendicular to the (*h*00) plane. This indicates that the stoichiometric thin film is primarily composed of particles with *h*00 orientation, whereas the non-stoichiometric thin films also contain particles without preferential orientation.

Figure 5 shows the compositional dependence of the lattice parameter $a_0$. The thin films fabricated without S-plasma supply, and the stoichiometric thin film fabricated with S-plasma at $T_{Sulfur}$ = 100 °C exhibited lattice parameters $a_0$ nearly identical to that of bulk SnS. In contrast, the non-stoichiometric thin films fabricated with an S-plasma supply exhibited expansion along the *a*-axis, which was particularly pronounced for S-rich compositions. Therefore, deviations from the stoichiometric composition appeared to induce tensile stress along the *a*-axis. Although expansion along the *a*-axis has been reported for S-poor SnS.[17] To the best of our knowledge, there are no reports on the lattice parameters of S-rich SnS. In this study, it was found that the compositional deviation toward the S-rich side had a stronger effect on the increase in the lattice parameter $a_0$. As mentioned earlier, the SnS thin film fabricated without S-plasma exhibited Raman peaks with the smallest FWHM (Figure 3(a)). In addition, among the thin films fabricated with S-plasma, the FWHMs tended to decrease as the composition approached stoichiometry. The peak positions of the Raman spectra indicate that thin films deviating from stoichiometry (i.e., those under tensile stress along the *a*-axis) exhibit greater shifts in the peaks at approximately 190 and 217 $cm^{-1}$ (Figure 3(d,e)). These findings suggest that deviations from stoichiometry increase the strain within the crystal and degrade the crystallinity of SnS thin films.

Figure 6 shows the results of rocking curve measurements (ω-scan) for the 400 peak of the thin films. The FWHM of the rocking curve of the stoichiometric thin film ($x_S$ = 0.5) was 2.8°, which is significantly narrower than those of the non-stoichiometric thin films. This result is consistent with the out-of-plane and in-plane XRD analyses, which indicated that the stoichiometric SnS thin film exhibited the highest degree of preferential orientation. The FWHMs of the SnS thin films grown on GaAs substrates by MBE and on MgO substrates by PLD have been reported to be 2.96° and 1.6°, respectively.[18,19] In the current study, despite being a self-oriented film deposited on a $SiO_2$ glass substrate, a high-quality thin film with a FWHM of the rocking curve comparable to that of epitaxial thin films was obtained, provided that a stoichiometric composition is achieved.

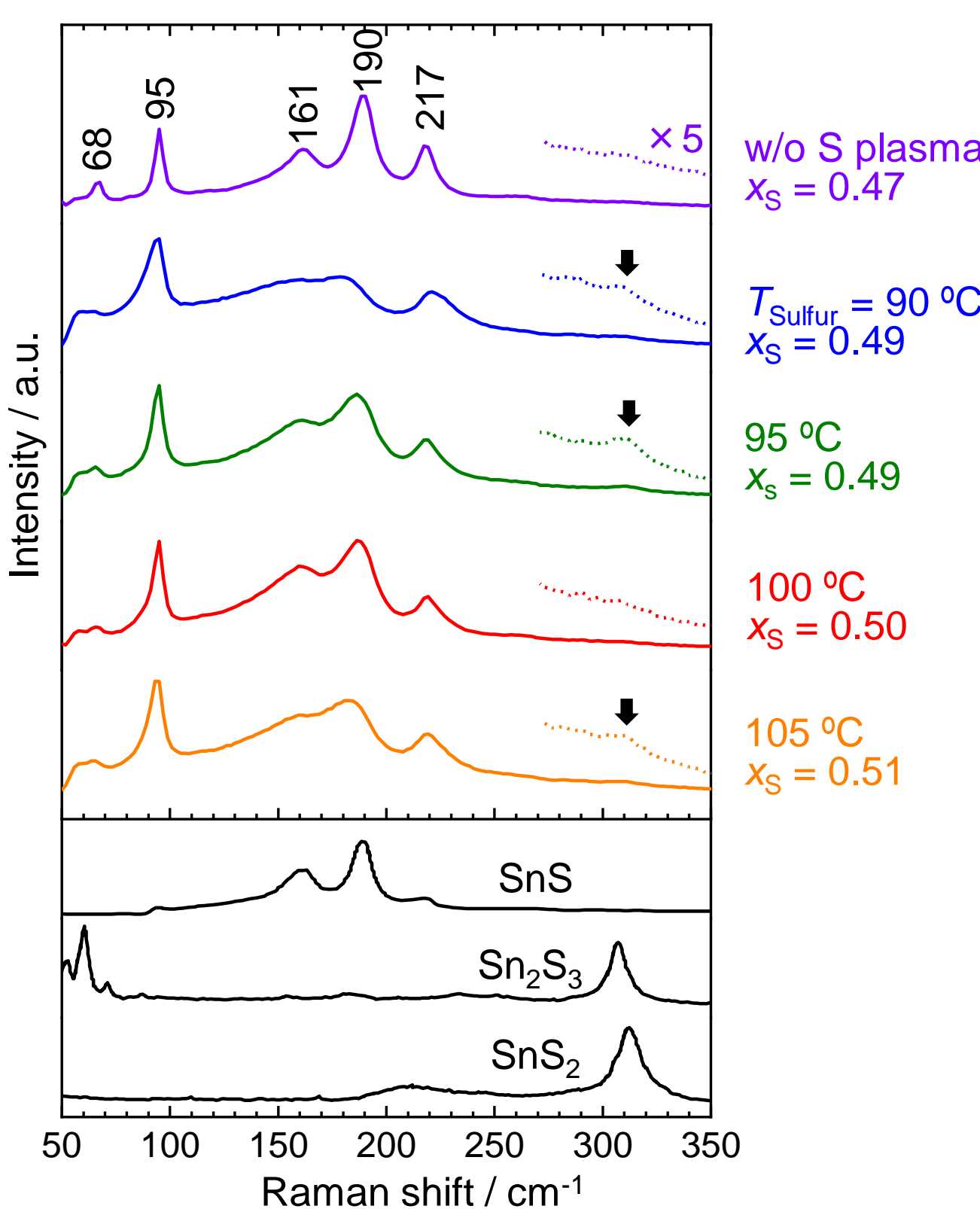

Figure 2. (a) Raman spectra of thin films fabricated with and without S-plasma supply, together with those of SnS, $Sn_2S_3$, and $SnS_2$ cited from literature[13,14]. Black arrows indicates peak corresponding to the main peak of $S_2S_3$ or $SnS_2$.

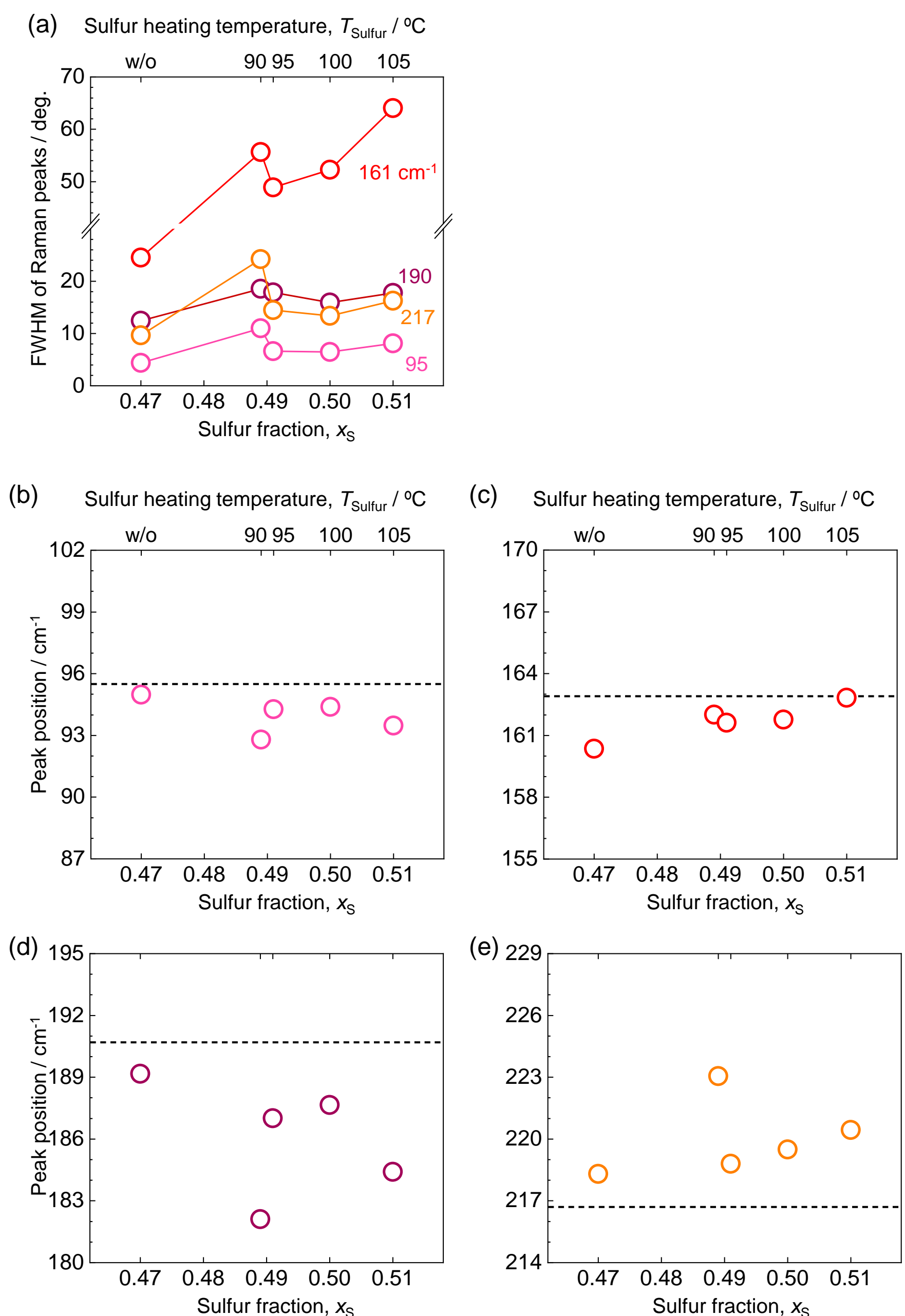

Figure 3. Compositional dependence of (a) FWHMs and (b–e) peak positions of representative Raman peaks observed at approximately 95, 161, 190, and 217 cm$^{-1}$. Dashed lines represent the literature values of these Raman peaks.[13]

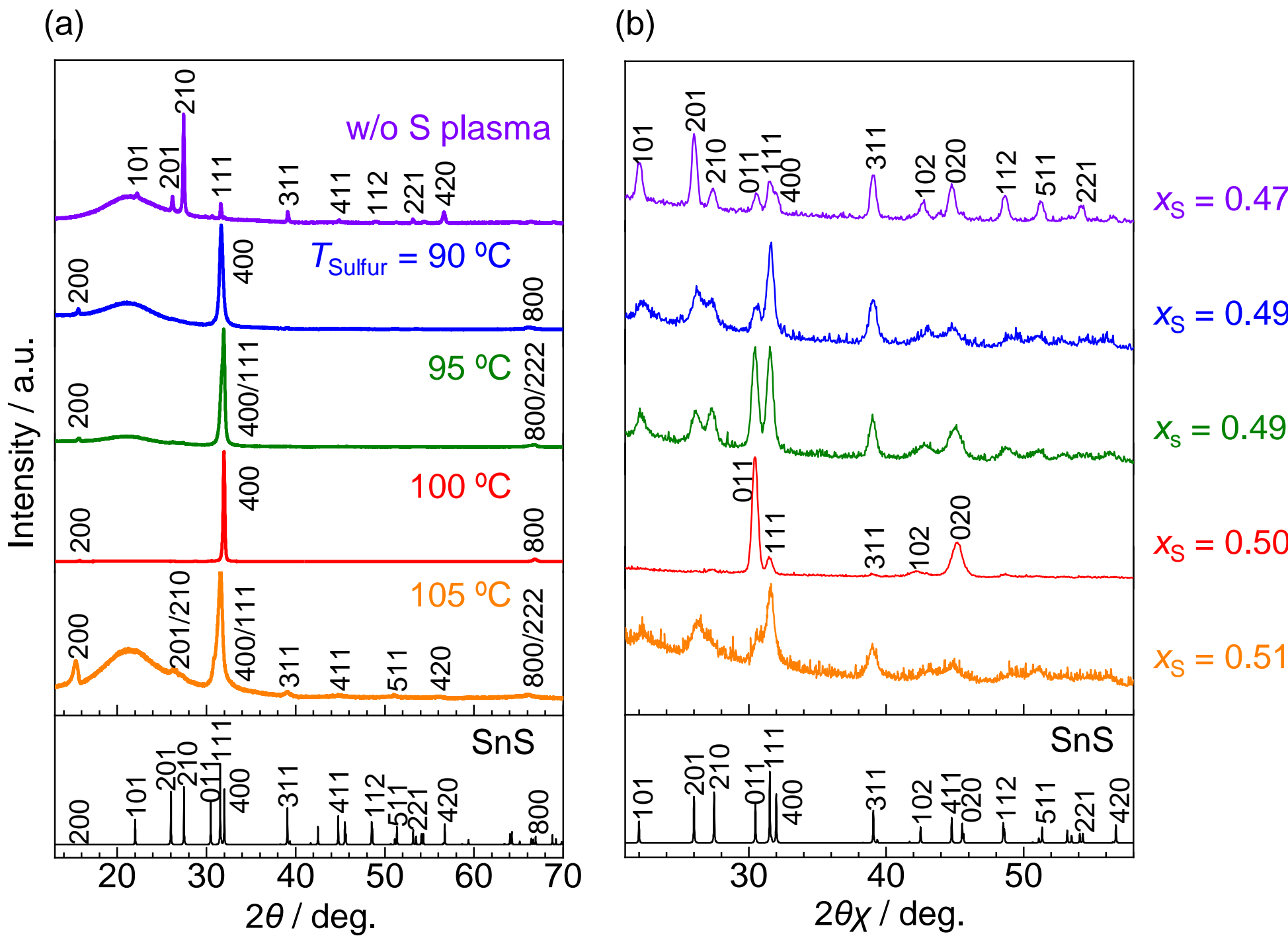

Figure 4. (a) Out-of-plane and (b) in-plane XRD profiles of thin films fabricated without and with S-plasma supply at various $T_{Sulfur}$, together with the powder pattern of α-SnS (ICSD#24376[20]).

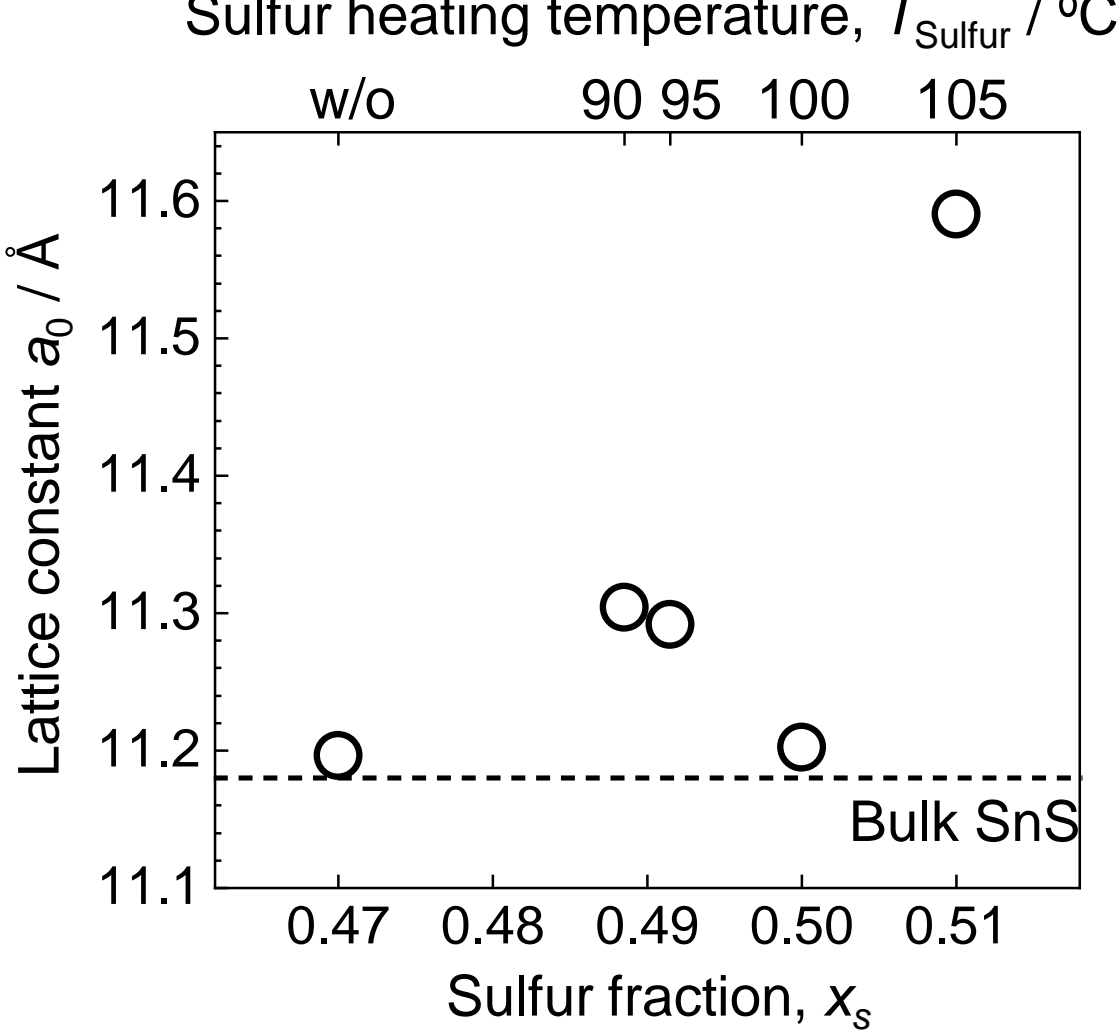

Figure 5. Compositional dependence of the lattice parameter $a_0$ of SnS thin films. The lattice parameter of bulk SnS is cited from literature (ICSD#24376[20]). The lattice parameters of the thin films fabricated with S-plasma were determined based on the 400 and 800 diffractions, while that of the thin film without S-plasma was determined based on the 201, 210, 111, 311, and 420 diffractions, all using out-of-plane XRD.

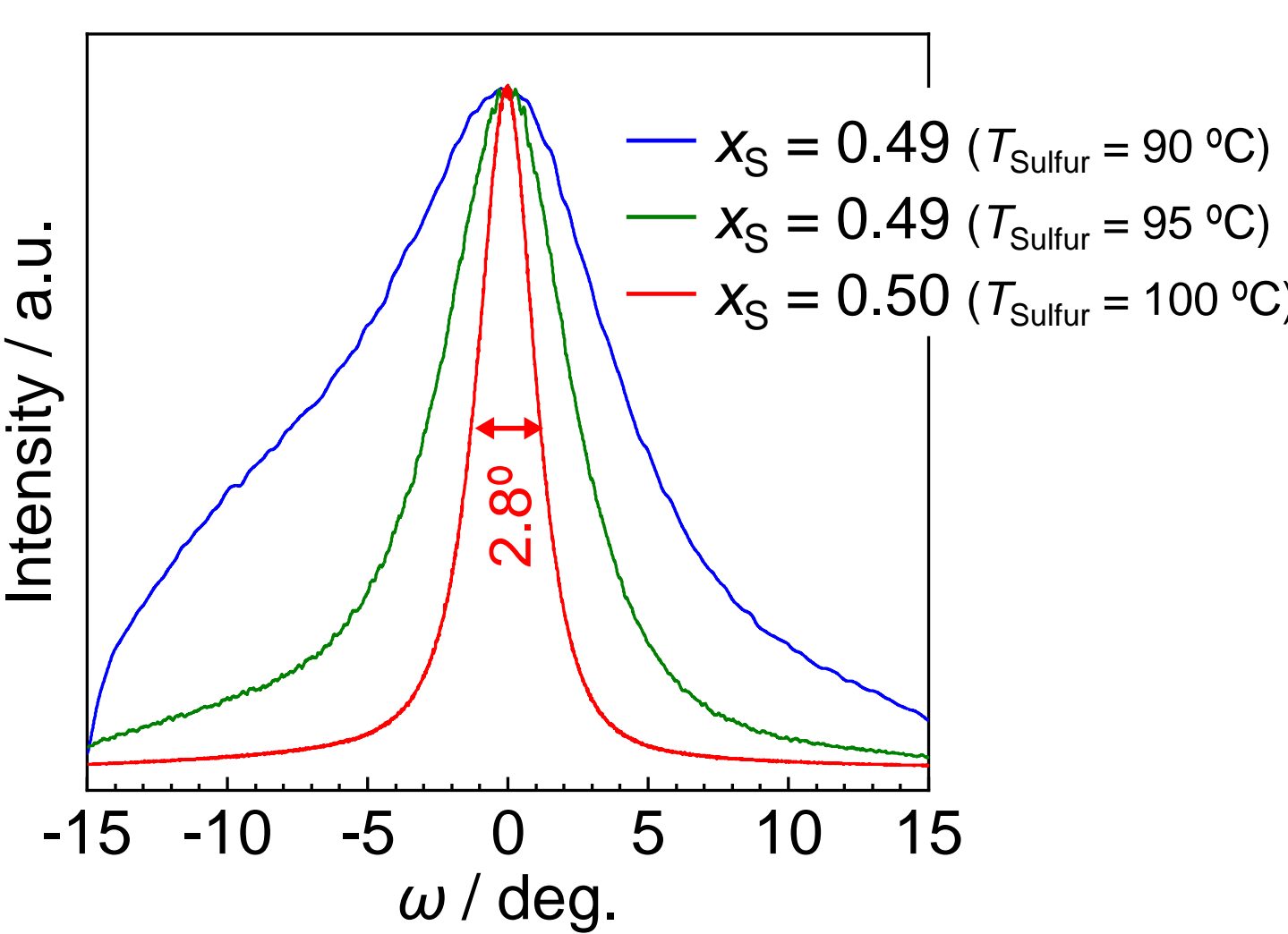

Figure 6. Rocking curve measurement (ω-scan) of SnS thin films with various compositions.

## 3.2 Influence of non-stoichiometry on thin-film morphology

Figure 7 shows surface and cross-sectional SEM images of the SnS thin films. The SnS thin film fabricated without S-plasma supply was composed of granular particles with diameters of approximately 200 nm. Non-stoichiometric SnS thin films fabricated with S-plasma supply at $T_{Sulfur}$ = 90, 95, and 105 °C exhibited disordered and sparsely deposited flake-shaped grains. In contrast, the stoichiometric SnS thin film ($T_{Sulfur}$ = 100 °C) exhibited a smooth surface and dense morphologies consisting of grains with diameters ranging from 200 to 500 nm.

Figure 8 shows schematic of the thin-film morphology, drawn based on the results of the SEM observations and XRD analysis described in the previous section. SnS, which has a layered crystal structure, exhibits a significantly lower surface energy on its van der Waals plane (i.e., the (100) plane) than the other planes.[21,22] This low surface energy serves as the driving force for the self-orientation of thin films, suggesting that the morphology of the stoichiometric thin films reflects the intrinsic characteristics of the layered structure. In contrast, thin films with flake-shaped grains have large surface areas of non-van der Waals planes. The reasons why non-stoichiometric SnS thin films do not grow with a minimum surface energy are currently unclear. As discussed in Section 3.1, non-stoichiometric SnS thin films contain trace amounts of $SnS_2$ or $Sn_2S_3$ (Figure 2), and these secondary phases tend to form on the surfaces of the SnS grains.[16] A study on the post-annealing of SnS thin films in a sulfur atmosphere reported that the smooth surface morphology drastically changed to a flake-like morphology associated with the formation of trace amounts of $SnS_2$.[23] These results suggest that the formation of secondary phases on the surfaces of the SnS grains modifies the surface energy of the van der Waals plane, thereby affecting the self-organization behavior.

Flake-like morphologies of SnS thin films have frequently been reported in studies on their fabrication using conventional RF magnetron sputtering, along with efforts to transform them into smooth morphologies. In all these studies, the substrate temperature and deposition pressure were identified as the key parameters determining their morphologies.[24,25] However, finding the deposition conditions that produces smooth and dense morphology is challenging, and moreover, these conditions are highly dependent on the deposition equipment, making it difficult to establish universal principles for determining parameters. The current study is the first to clearly demonstrates that the underlying cause of such morphological changes is slight non-stoichiometry, either on the S-poor or S-rich side, and that achieving stoichiometric composition enables the formation of a smooth and dense morphology. Considering that the sparse morphology of thin films has negative effects when used in devices, such as low shunt resistance in solar cell applications,[26] achieving a stoichiometric composition is desirable from a morphological perspective.

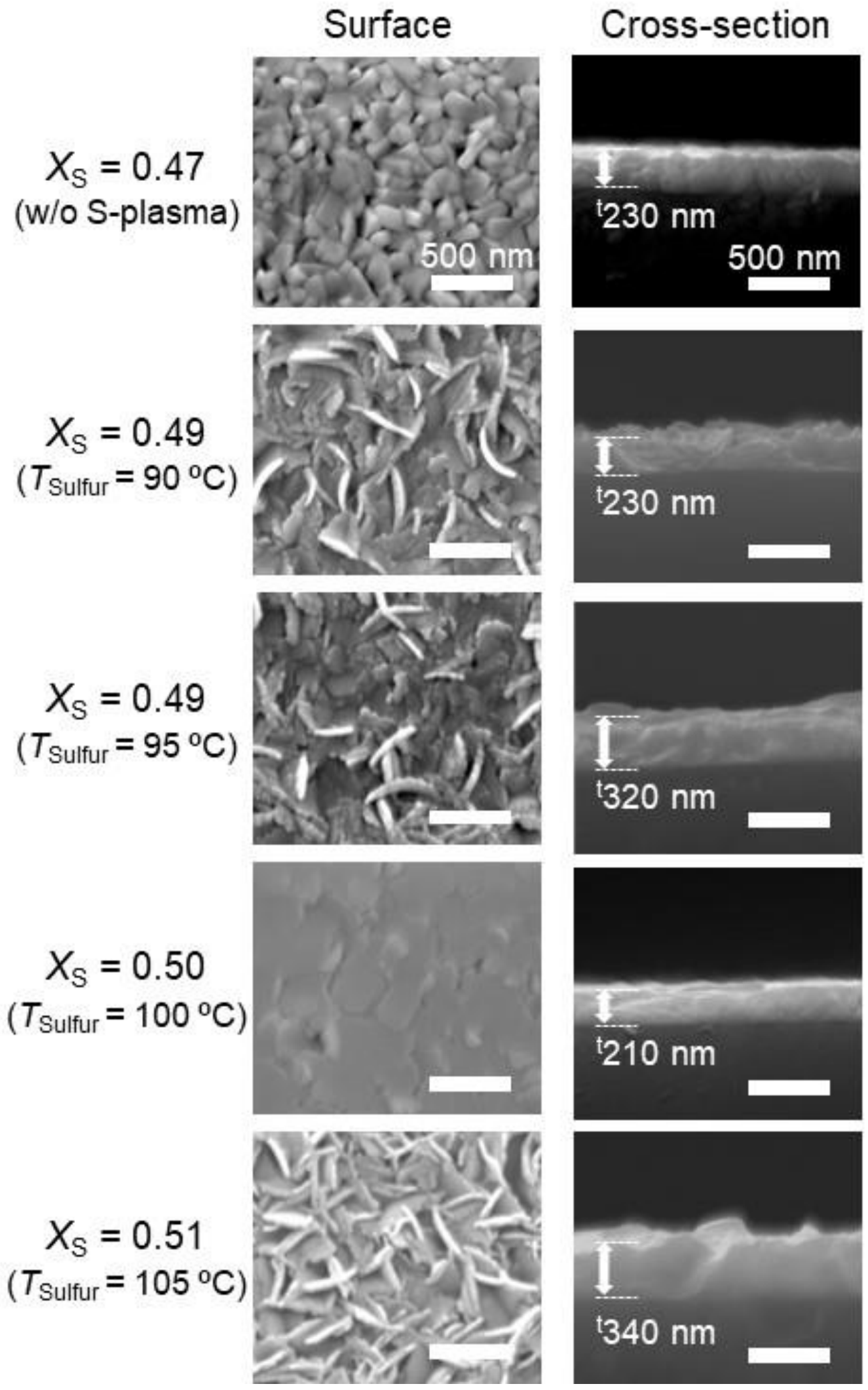

Figure 7. (a) Surface and (b) cross-sectional SEM images of SnS thin films fabricated with various S-plasma supply conditions.

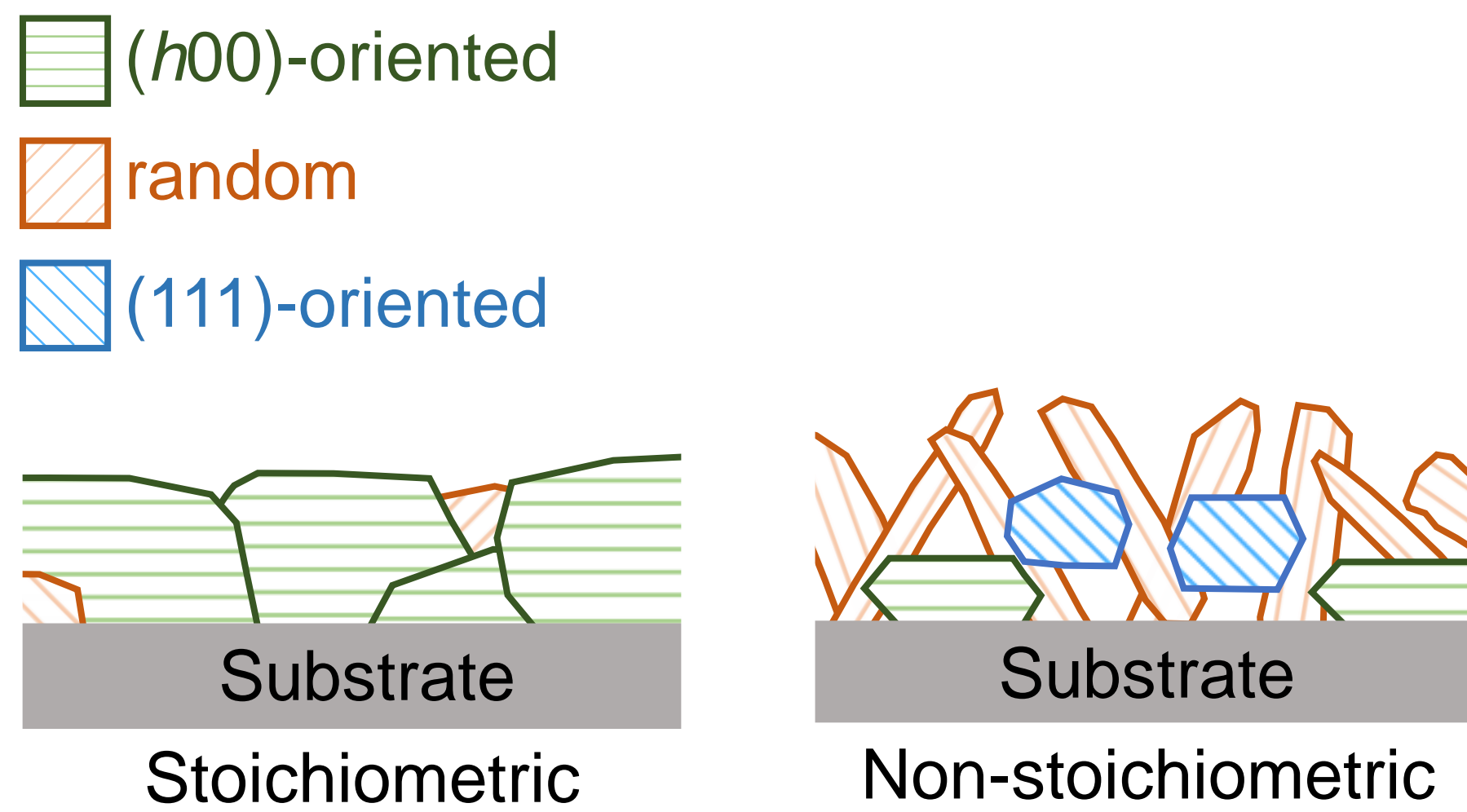

Figure 8. Schematic of the grain morphology and preferential orientation of the SnS thin films influenced by compositional deviations. The colors of the particles in the figure represent their preferential orientation.

### 3.3 Influence of non-stoichiometry on intrinsic defects and electrical properties

The SnS thin films fabricated in this study exhibited positive Hall coefficients, indicating their p-type conduction. Additionally, the Seebeck coefficient near room temperature for all thin films ranged from +60 to +730 μV/K (see details in Table S1 in the supporting information), further supporting their p-type behavior. Figure 9(a,b) show the compositional dependence of the Hall mobility and carrier density of the SnS thin films, respectively. 1While the Hall mobilities of the SnS thin films with S-poor compositions were approximately 2 $cm^2 V^{-1} s^{-1}$, it increased significantly to 15 $cm^2 V^{-1} s^{-1}$ in the stoichiometric composition and then abruptly decreased to 0.2 $cm^2 V^{-1} s^{-1}$ as the composition entered the S-rich region. The carrier densities in the S-poor region were relatively low, approximately $10^{15}$–$10^{16}$ $cm^{-3}$, while it increased significantly to the order of $10^{19}$ $cm^{-3}$ in the S-rich region. Several studies have reported that carrier scattering in SnS thin films is dominated by grain boundary scattering.[18,27,28] The stoichiometric SnS thin film exhibited a denser morphology than those with non-stoichiometric compositions (Figure 8), suggesting that reduced grain boundary scattering contributed to the higher mobility of the stoichiometric thin film. It has also been reported that impurity-doped SnS with a carrier density of approximately $10^{19}$ $cm^{-3}$ exhibits predominant ionized impurity scattering.[29] Therefore, the low Hall mobility of SnS thin films with an S-rich composition is likely due not only to morphological factors but also to ionized impurity scattering within the particles.

Several first-principles studies have reached a consensus that, based on the calculated formation enthalpies of various defects under equilibrium conditions, the most readily formed defect at S-rich condition is Sn vacancy ($V_{Sn}$).[7–9] As mentioned above, as sputtering is a non-equilibrium deposition technique, the defect species induced by non-stoichiometry could differ from those predicted by first-principles calculations. However, for the purposes of this discussion, we assume that $V_{Sn}$ is also the most readily formed defect species during sputtering. Because $V_{Sn}$ is an acceptor-type defect, it could be the origin of the high carrier density in S-rich SnS thin films. Figure 10 shows the relationship between the composition and density of $V_{Sn}$ ([$V_{Sn}$]), assuming that the deviation from stoichiometry toward the S-rich side leads exclusively to the formation of $V_{Sn}$. Based on this assumed relationship, [$V_{Sn}$] would be approximately $1 \times 10^{21}$ $cm^{-3}$ in the case of $x_S = 0.51$ (corresponding to the chemical formula of $Sn_{0.96}S$). The actual carrier density was $2 \times 10^{19}$ $cm^{-3}$, which indicates that 1% of $V_{Sn}$ are activated and contribute to the generation of carrier holes assuming that $V_{Sn}$ is a divalent acceptor. This activation rate of 1% is close to that of the dopants in SnS, for instance, 0.5% and 1.5% for Cl- and Br-doped n-type SnS[30], respectively, and 2% and 3% for Na- and Ag-doped p-type SnS[31,32], respectively. Therefore, the assumption that the high carrier density in the S-rich SnS thin films obtained in this study was caused by $V_{Sn}$ originating from deviations from stoichiometry is reasonable. A significant increase in the carrier density due to intentional compositional deviation toward the S-rich side has also been reported for SnS nanoflakes.[33]

First-principles calculations under equilibrium conditions indicated that the S vacancy ($V_S$), which acts as a donor-type defect with a Fermi level close to the valence band maximum, easily forms under S-poor conditions. Along with the formation of $V_S$, Sn migrates to the $V_S$, generating Sn antisites ($Sn_S$) and $V_{Sn}$, both of which act as acceptor-type defects.[7–9] Assuming that the deviation from stoichiometry toward the S-poor exclusively leads to the formation of these defects, as in the S-rich condition discussed above, $[V_S]$ + $([Sn_S] + [V_{Sn}])/2$ would be approximately 1 × $10^{21}$ $cm^{-3}$ at $x_S$ = 0.49 (corresponding to the chemical formula of $SnS_{0.96}$) (Figure 10). It is interesting that the carrier densities of S-poor SnS thin films remained nearly constant at $10^{15}$–$10^{16}$ $cm^{-3}$, regardless of the compositions. These relatively low carrier densities can be attributed to self-compensation between the acceptor-type defects ($Sn_S$ and $V_{Sn}$) and donor-type defects ($V_S$). In other words, the difference in the density of ionized acceptor- and donor-type defects in the S-poor region (i.e., $[Sn_S^-]$ + $[V_{Sn}^{2-}]$ – $[V_S^{2+}]$) is supposed to be approximately $10^{16}$ $cm^{-3}$, regardless of the composition. This value is consistent with the reported effective acceptor density ($N_{A,eff} = N_A - N_D$) of approximately $10^{16}$ $cm^{-3}$ in SnS, as determined by Mott–Schottky (capacitance–voltage) measurements.[34–36] Furthermore, photoelectron spectroscopy analysis indicated that S-poor SnS thin films exhibited Fermi level pinning due to defects associated with sulfur deficiency[37,38], consistent with the constant carrier density observed for S-poor SnS in the current study.

The highest defect densities under equilibrium conditions in SnS were estimated by first-principles calculations to be $10^{18}$ and $10^{15}$ $cm^{-3}$ for the S-rich and S-poor limits, respectively,[7,9] which are a few orders of magnitude smaller than the densities estimated based on the deviation from stoichiometry described above. In the first-principles calculations, the S-rich and S-poor limits refer to conditions in which SnS is in equilibrium with other competing phases, such as elemental S, $SnS_2$, $Sn_2S_3$, and metallic Sn. During sputtering, particles in a non-equilibrium state are rapidly quenched on the substrate, often resulting in the formation of metastable states. Therefore, it is important to note that the conditions employed in first-principles calculations (S-rich and S-poor conditions) fundamentally differ from the actual compositional deviation towards S-rich and S-poor in thin films. Despite the significant non-stoichiometry in some thin films obtained in the current study, their crystal structures were the same as the original SnS. The non-equilibrium nature of sputtering is likely responsible for producing higher defect densities in the thin films than those predicted by first-principles calculations.

To investigate the effect of the deviation from stoichiometry on the origin of carriers, the temperature dependence of the electrical conductivity was measured (Figure S4 of the Supplementary Material). Figure 9(c) shows the energy difference between the acceptor level with respect to the valence band maximum ($E_A - E_{VB}$) determined from these measurements. On the S-poor side, this value remains almost constant at approximately 320 meV. As the composition deviated toward the S-rich side, the acceptor level became significantly shallower, reaching $E_A - E_{VB}$ = 160 meV at $x_S$ = 0.51. This indicated that the defect levels responsible for the carriers changed as the composition shifted across the stoichiometric boundary. This is consistent with the first-principles calculation results, indicating that SnS at the S-poor and S-rich conditions tend to form different defect species. Figure 11 shows the energy levels for representative defects ($Sn_S$, $V_{Sn}$, and $V_S$) based on the first-principles calculations.[7–9] The defect levels reported in various first-principles calculation studies do not exactly match due to differences in the choice of functional, the size of the supercell, and the constraints applied to the geometrical relaxation of the interlayer distance.[10] Nevertheless, it is generally accepted that $Sn_S$ and $V_S$ form relatively deep acceptor and donor levels, respectively, while $V_{Sn}$ forms shallow acceptor levels. The calculated energy of the lowest acceptor level of $Sn_S$ ($E_A - E_{VB} \approx$ 250–700 meV) is higher than that of the highest acceptor level of $V_{Sn}$ ($E_A - E_{VB} \approx$ 150–220 meV). This is consistent with the experimental observation that $E_A - E_V$ significantly decreases when the composition varied from S-poor to S-rich.

While the value of $E_A - E_{VB}$ of the stoichiometric SnS thin film was between those of $x_S = 0.49$ and 0.51, its carrier density was similar to that of S-poor thin films. These results indicate that self-compensation involving high-density donors and acceptors, as observed in S-poor thin films, does not occur in the stoichiometric thin film. The formation of high-density acceptors, as observed in the S-rich thin film, was also absent, suggesting a low defect concentration in the stoichiometric thin film. Figure 12 shows the absorption spectrum of the stoichiometric SnS thin film ($x_S = 0.5$) determined by spectroscopic ellipsometry (details of the analysis are provided in Figure S5 of the Supplementary Material). In addition to the strong absorption corresponding to the fundamental absorption of SnS, a small dip just above the band gap was observed, which is consistent with the previously reported absorption spectrum obtained by transmittance measurements[39]. When SnS thin films contain high-density sulfur deficiencies, broad absorption corresponding to the in-gap states formed by the associated defects is observed at energies below the band gap[37,40] , which was not observed in the current study. In general, spectroscopic ellipsometry can detect high-density defects (>$10^{18}$ $cm^{-3}$) that affect optical properties.[41,42] The defect density in the stoichiometric SnS thin film fabricated in the current study is not considered high, which is consistent with the electrical measurement results.

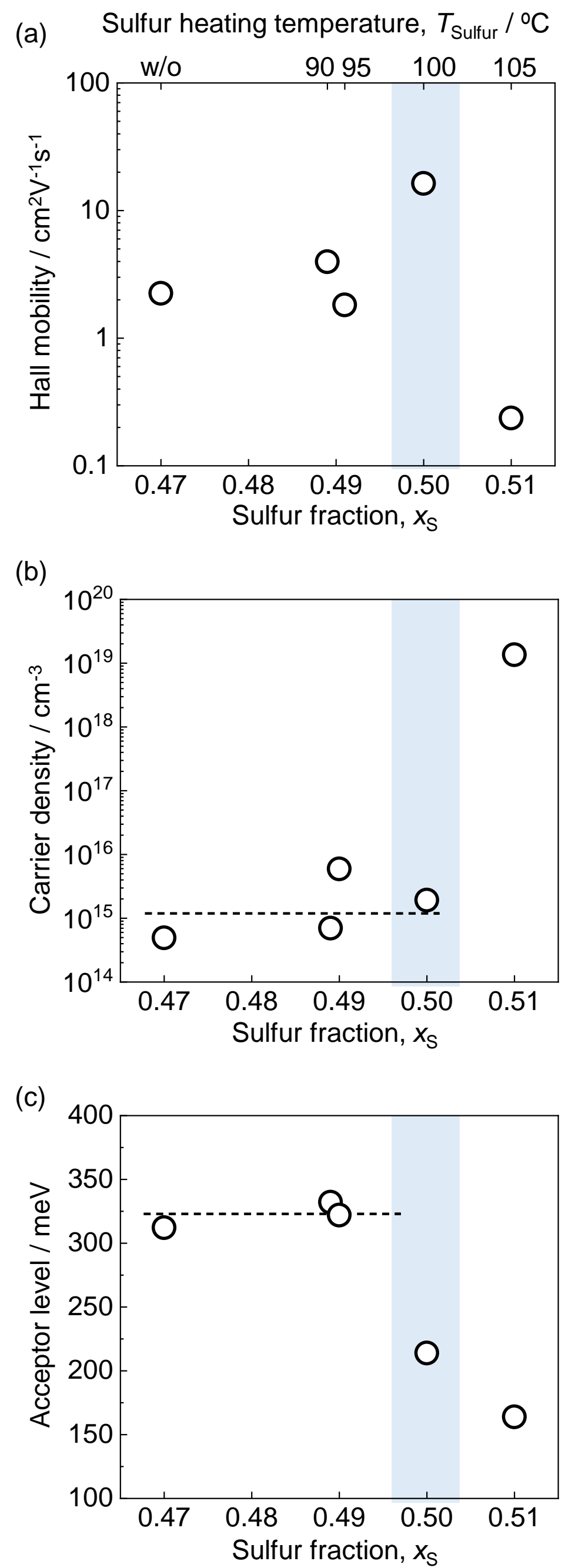

Figure 9. Compositional dependence of (a) Hall mobility and (b) carrier density of SnS thin films deposited without and with S-plasma supply determined by Hall measurements. (c) Acceptor level of the SnS thin films determined by the temperature dependence of the electrical conductivity.

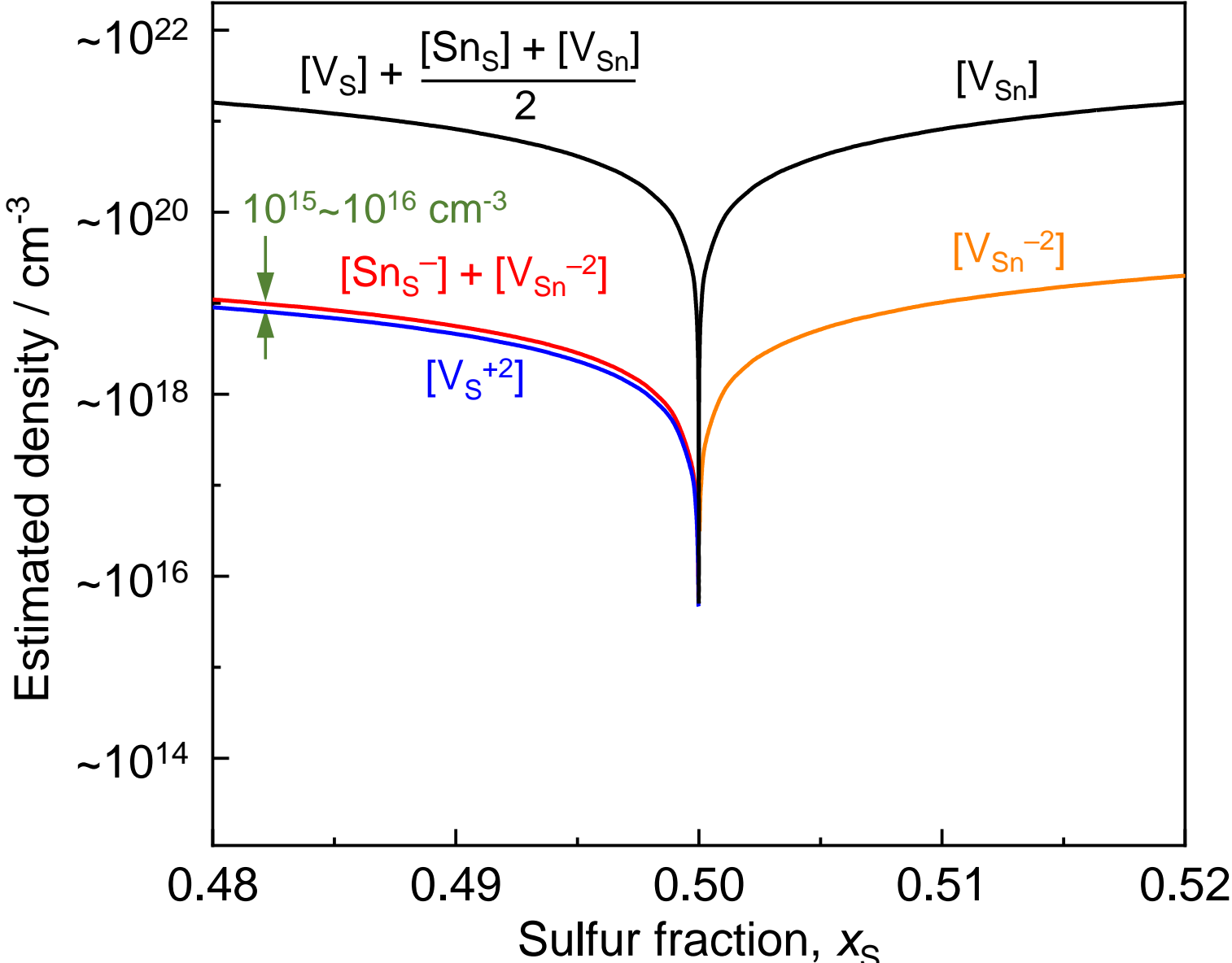

Figure 10. Schematics of the correlation between deviation from stoichiometry and intrinsic defect densities. Defect species labeled with charges are ionized.

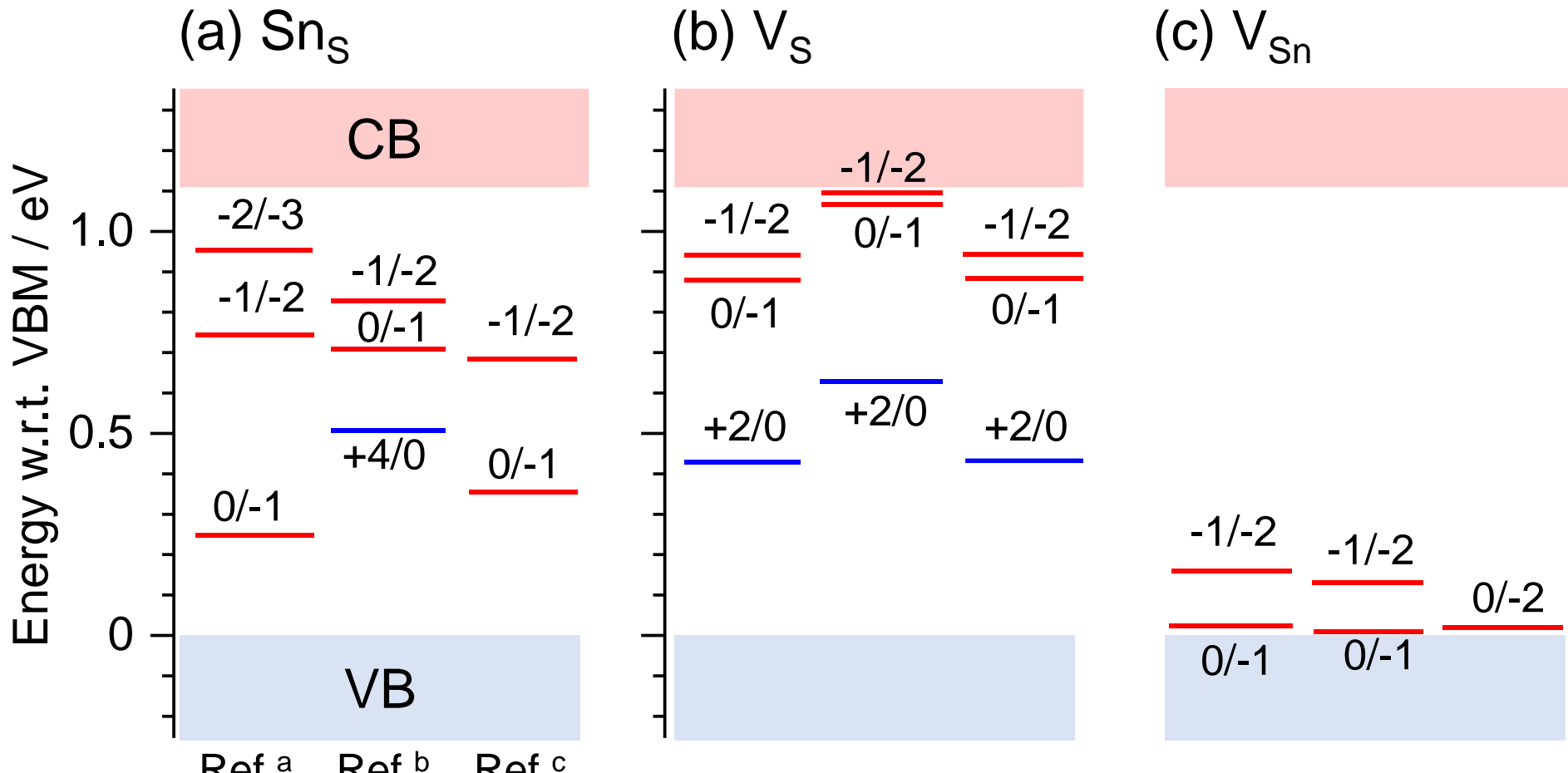

Figure 11. Calculated energy levels of representative intrinsic defects ($Sn_S$, $V_{Sn}$, and $V_S$) with respect to the valence band (VB) and conduction band (CB). Values are cited from [a]Ref.[7], [b]Ref[8], and [c]Ref.[9]. Energy levels in red and blue correspond to acceptor- and donor-type levels, respectively.

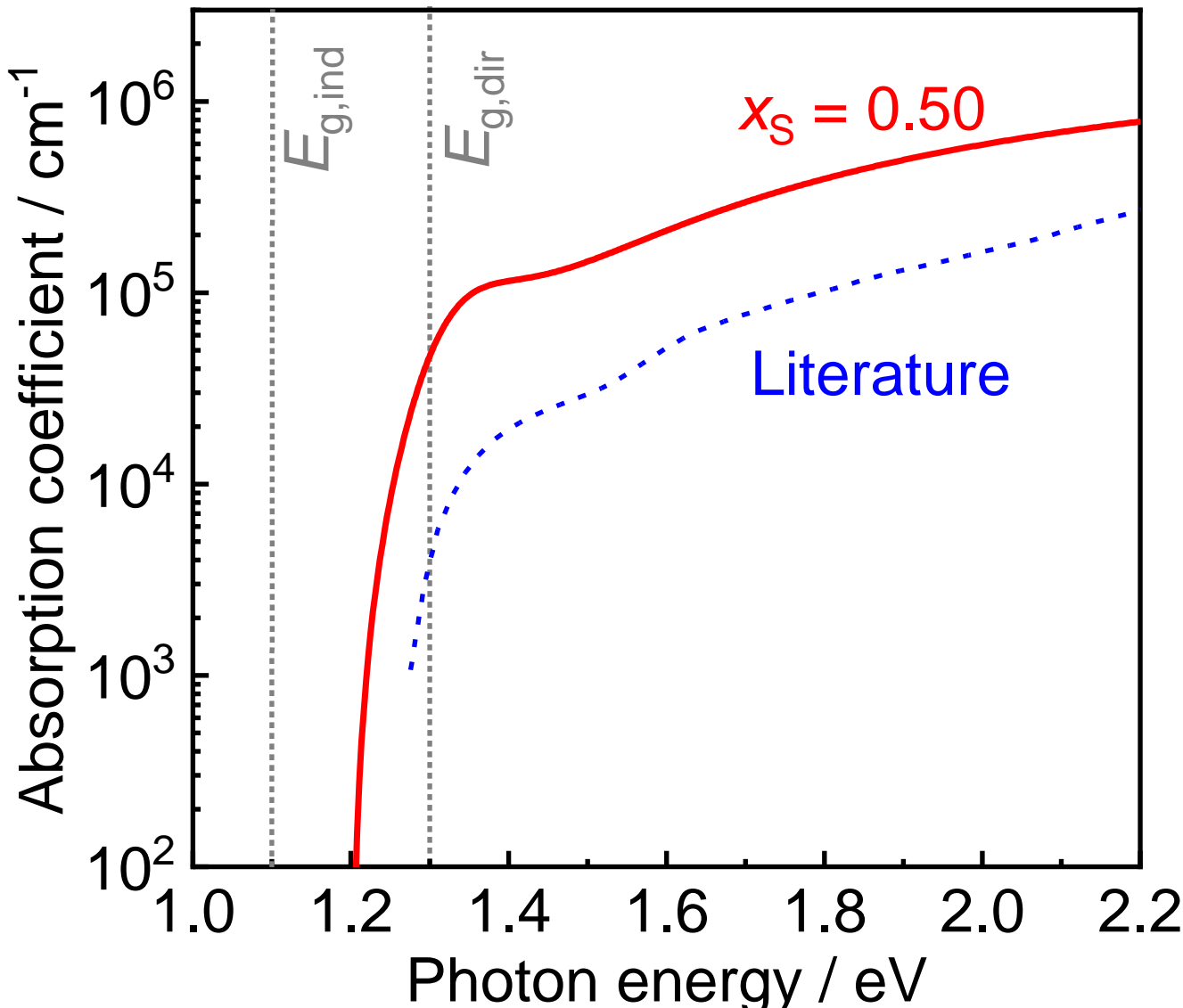

Figure 12. Absorption coefficient obtained from spectroscopic ellipsometry of the stoichiometric SnS thin film ($T_{Sulfur}$ = 100 °C). The blue spectrum was taken from the literature.[39]

## 3.4 Importance of sulfur content control in SnS

In previous studies on SnS thin films, their compositions were determined using X-ray photoelectron spectroscopy (XPS), energy-dispersive X-ray spectroscopy (EDX), EMPA, and X-ray fluorescence (XRF). Although these measurements provide reliable relative comparisons within a series of thin films, assessing the absolute compositions requires considerable caution and is often inaccurate. For example, XRF and EPMA can easily yield discrepancies of several percent in composition, even when measuring the same sample.[43] In the current study, the thin film with stoichiometric composition (xS = 0.5) exhibits a distinctive electrical property, thereby validating the reliability of the assessment of stoichiometric composition. Conventionally reported undoped SnS thin films have all exhibited relatively low carrier densities within a narrow range of approximately $10^{15}$–$10^{16}$ $cm^{-3}$.[44,45] The reported compositions of these thin films include not only S-poor ones ($0.45 \leq x_S < 0.5$)[37,46–49] but also S-rich ones ($0.5 < x_S \leq 0.52$[50–53]). Considering the difficulty in evaluating absolute compositions and the results of the current study, which shows a sharp increase in carrier density for S-rich compositions, it is strongly suggested that SnS thin films reported as both S-poor and S-rich may in fact actually have S-poor compositions.

Owing to the extremely high vapor pressure of sulfur, sulfur deficiency is commonly observed in the deposition of various sulfide semiconductors, as well as in SnS. Sulfur deficiency in sulfides results in $V_S$, which typically functions as a donor-type defect. For instance, CdS and $SnS_2$ with relatively narrow band gaps of 2.3 and 1.9 eV, respectively, which are slightly larger than that of SnS, intrinsically exhibit n-type conduction due to donor-type defect $V_S$. In these simple sulfides, the compositional deviation from stoichiometry toward the S-poor side is directly reflected in an increase in carrier density and is readily detectable.[54,55] In contrast, SnS exhibits p-type conductivity even in S-poor composition, and its carrier density is insensitive to the deviation from stoichiometry toward the S-poor side. The reason why SnS exhibits a behavior different from that of CdS and $SnS_2$ can be qualitatively understood, as shown in Figure 13, where the effect of the matrix in each material on the formation enthalpy of defects is neglected for simplicity. In CdS and $SnS_2$, metal cation vacancies act as acceptor-type defects, and the formation enthalpy of donor-type defects, including $V_S$, is low, leading to n-type conductivity.[10,56] The valence band of SnS is mainly composed of the Sn 5s orbital,[57] whereas those of CdS and $SnS_2$ are mainly composed of S 3p orbitals. Because the energy of the Sn 5s orbital is higher than that of the S 3p orbital, the valence band of SnS is pushed upward compared to those of CdS and $SnS_2$.[58] As a result, the formation enthalpy of intrinsic acceptors in SnS is relatively low. Therefore, SnS exhibits p-type conduction owing to self-compensation, even in the presence of donor-type $V_S$. Similarly, in SnO, where the Sn 5s orbitals contribute to the valence band, it has been reported that acceptor-type $V_{Sn}$ and donor-type $V_O$ compete, generally resulting in p-type conduction.[59] Nevertheless, due to the narrow band gap of SnO (0.7 eV), it can also exhibit n-type conduction under O-poor conduction due to donor-type $V_O$.[60]

In SnS thin films, non-stoichiometry toward the S-rich side significantly affects the carrier density and conductivity, whereas non-stoichiometry toward the S-poor side has almost no impact on these electrical properties. This has led to the conventional perception that the extent of sulfur deficiency in SnS thin films requires minimal consideration. It should be noted that a significant sulfur deficiency can lead to the generation of a large number of defects, which in turn can cause Fermi-level pinning to degrade photovoltaic performance.[38] Therefore, it is important to understand that even if the carrier density in SnS thin films is comparable to the low acceptor density suitable for solar cells ($<10^{16}$ $cm^{-3}$ [61]), this does not immediately indicate that they are suitable for application in solar cell devices.

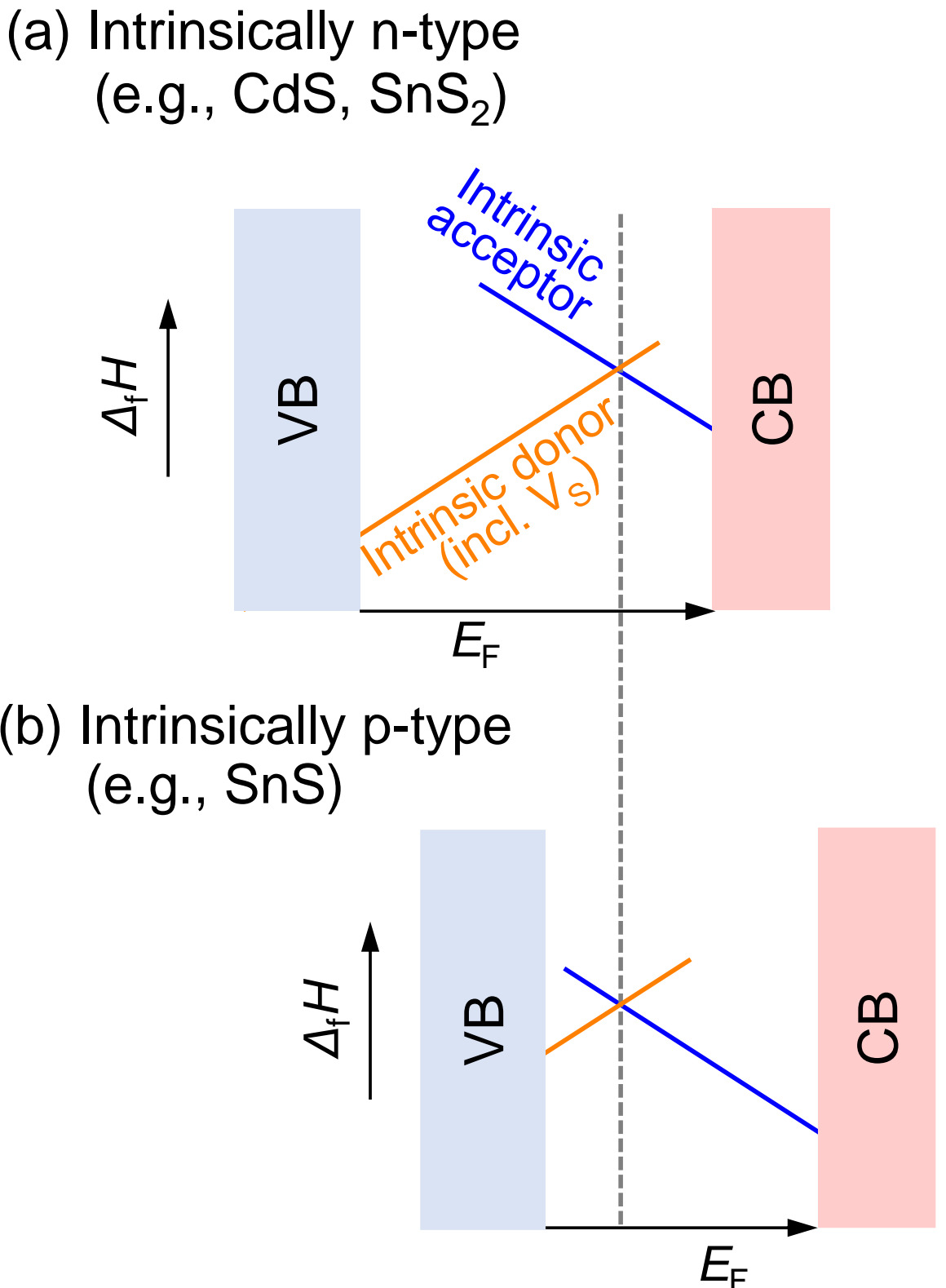

Figure 13. Schematic of the relationship of the formation enthalpy ($\Delta_f$H) of competing intrinsic donor- and acceptor-type defects. Note that this extremely simplified model ignores the effect of the matrix of each material on defect formation and only considers the band alignment for qualitative understanding.

## 4. Conclusion

This study investigated how deviations in composition near stoichiometry affect the properties of SnS thin films by utilizing a unique approach of sputtering SnS with a superimposed supply of S-plasma to realize precise control of sulfur content. Thin films fabricated under non-equilibrium conditions exhibited a significant number of defects owing to non-stoichiometry. In particular, in S-rich compositions, the carrier density was found to increase significantly, whereas the hole mobility decreased, which is attributed to the intrinsic acceptor-type defects $V_{Sn}$. In contrast, in S-poor compositions, the carrier density and effective defect density were found to be insensitive to non-stoichiometry because intrinsic acceptor-type defects ($Sn_S$ and $V_{Sn}$) and donor-type defects ($V_S$) mutually compensate. This suggests that the high-density defects present in S-poor SnS thin films have previously been overlooked or given insufficient consideration. Furthermore, it was demonstrated that the underlying origin determining the preferential orientation and morphology of SnS thin films lies in non-stoichiometry, and that achieving stoichiometry enables higher orientation as well as smoother and denser morphologies suitable for device applications. These results indicate that precise control over the composition of SnS thin films is crucial for optimizing their crystallinity and electrical properties, including carrier density and carrier transport properties, and, ultimately, the electronic states at the interfaces with other semiconductors when applied in devices.

An additional yet significant aspect to emphasize is the advantage of the sputtering method employed in this study as a versatile approach for depositing various sulfides. The formation of sulfur deficiencies is a common phenomenon in the fabrication process of sulfide thin films owing to the high vapor pressure of sulfur. For example, sulfur deficiency can cause issues such as reduced optical transparency in CdS thin films[62] and degradation of the switching performance in transistors made from transition metal dichalcogenides such as $MoS_2$ and $WS_2$.[63] Two main approaches have been employed to compensate for sulfur deficiency. One approach involves deposition by sputtering or PLD in an $H_2S$-containing atmosphere.[40,64] The other approach is post-annealing in an atmosphere with elevated sulfur partial pressure using $H_2S$ or sulfur powder.[65,66] However, $H_2S$ is highly toxic and undesirable in industrial processes. More fundamentally, annealing with $H_2S$ inevitably results in the incorporation of significant amounts of hydrogen impurities into the material, which alters its properties.[67] Moreover, post-annealing presents challenges in achieving the uniform diffusion of sulfur from the thin-film surface throughout the film. In particular, for compounds containing metal ions capable of multiple oxidation states, determining the optimal conditions for uniform sulfurization is challenging. As exemplified by SnS, increasing the sulfur partial pressure can easily oxidize $Sn^{2+}$ to $Sn^{4+}$, leading to the formation of $SnS_2$ and $Sn_2S_3$.[68] The deposition method employed in this study, which combines sputtering with a S-plasma supply, effectively addresses the aforementioned issues by providing impurity-free sulfur during thin-film deposition. This approach is expected to serve as a versatile process for fabricating high-quality sulfide thin films with reduced sulfur deficiencies.

## Associated Content

Supplementary Material:

Evaporation amount of sulfur and plasma emissions, out-of-plane XRD analysis, Seebeck coefficient measurements, temperature dependence of electrical conductivity, and spectroscopic ellipsometry analysis.

## Author Contributions

**Taichi Nogami**: Investigation (lead), Visualization (equal), Writing – original draft (equal). **Issei Suzuki**: Conceptualization (lead), Investigation (supporting), Visualization (equal), Funding acquisition (lead), Supervision (equal), Writing – original draft (equal). **Daiki Motai**: Investigation (supporting). **Hiroshi Tanimura**: Investigation (supporting). **Tetsu Ichitsubo**: Investigation (supporting). **Takahisa Omata**: Supervision (equal), Funding acquisition (supporting), Writing – review and editing (lead).

## Acknowledgments

This work was partly supported by a Grant-in-Aid for Scientific Research (B) (Grant No. 21H01613), the Research Program of "Five-star Alliance" in "NJRC Mater. & Dev.", and the Tsukuba Innovative Arena (TIA) "Kakehashi" collaborative research program. T. Nogami was financially supported by JST SPRING, Grant Number JPMJSP2114.

### Author Information

Taichi Nogami

Issei Suzuki https://orcid.org/0000-0002-0869-2713

Daiki Motai https://orcid.org/0009-0007-5210-0053

Hiroshi Tanimura https://orcid.org/0000-0002-7343-1966

Tetsu Ichitsubo https://orcid.org/0000-0002-1127-3034

Takahisa Omata https://orcid.org/0000-0002-6034-4935

Highlight image

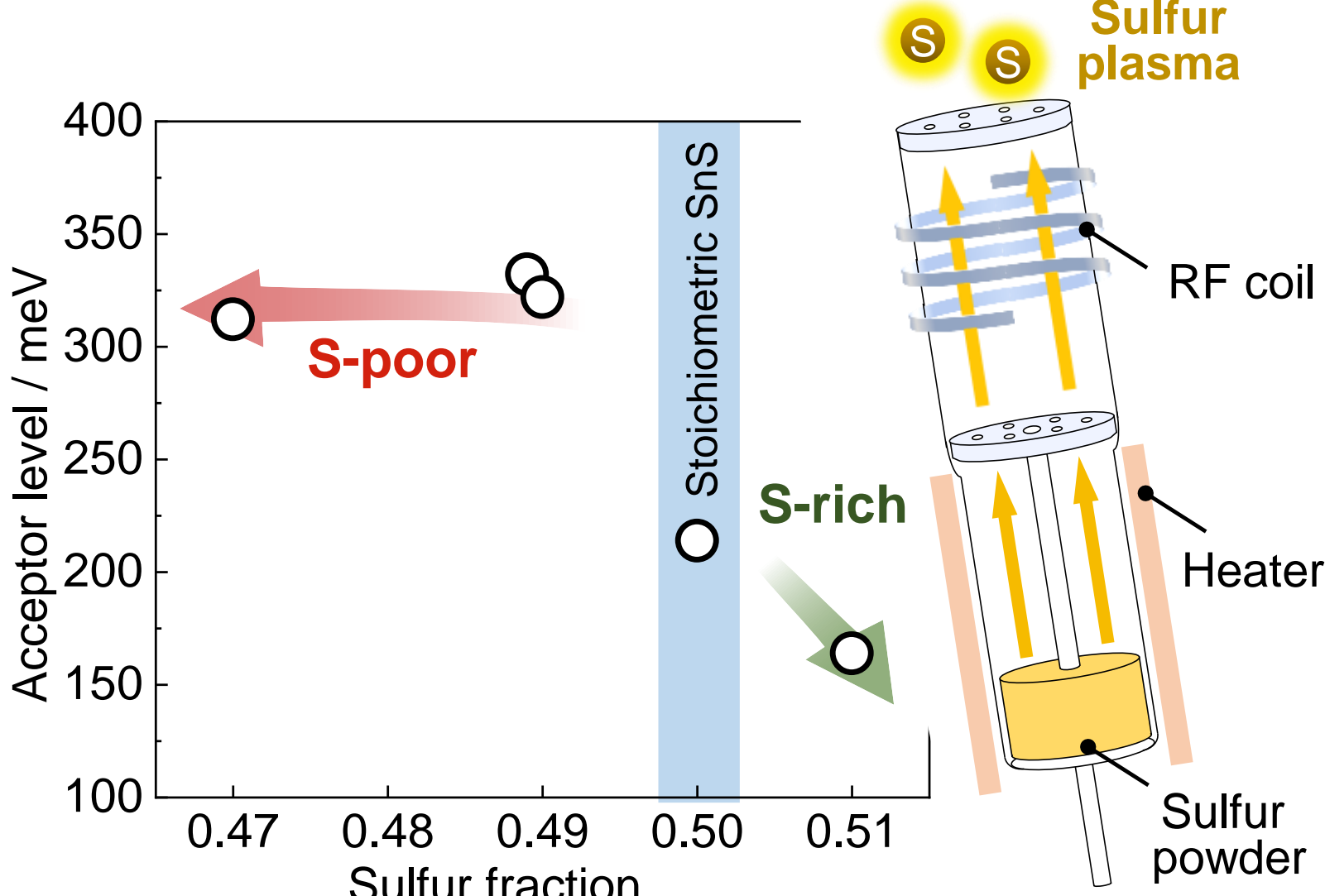

**Supplementary Material for**

# Non-stoichiometry in SnS: How it affects thin-film morphology and electrical properties

Taichi Nogami[1], Issei Suzuki*,[1], Daiki Motai[1],
Hiroshi Tanimura[2], Tetsu Ichitsubo[2], Takahisa Omata[1]

1. Institute of Multidisciplinary Research for Advanced Materials, Tohoku University, Sendai 980-8577, Japan

2. Institute for Materials Research, Tohoku University, Sendai 980-8577, Japan

*Corresponding author: issei.suzuki@tohoku.ac.jp

## Section S1. Evaporation amount of sulfur and plasma emission

When the $T_{Sulfur}$ was set to 100 °C, the sulfur powder inside the S-plasma source decreased by 10–20 mg over a 3 h operation. For example, assuming that a 200-nm-thick SnS thin film was deposited on a 40 × 40 $mm^2$ substrate for 3 h, the amount of sulfur in the film was approximately 0.35 mg. In this study, the sulfur content increased by approximately 3% when deposited with S-plasma supply at $T_{Sulfur}$ = 100 °C compared to deposition without S-plasma. This corresponds to an increase of approximately 0.01 mg in the sulfur content of the SnS thin films. Therefore, 0.05–0.1% of the sulfur supplied from the S-plasma source was incorporated into the thin film, whereas the remaining majority was either deposited outside the substrate in the chamber or evacuated from the chamber.

Figure S1 shows the emission spectrum of the S-plasma measured by introducing an optical fiber at the bottom of the S-plasma source and using a spectrometer (PMA-11 C7473, Hamamatsu Photonics K.K., Japan). In addition to the strong emission from the Ar plasma observed at 700–900 nm, numerous peaks with relatively low intensities were observed at 250–650 nm. These peaks correspond to the plasma emissions of $S_2$–$S_8$.[S1,S2] As $T_{Sulfur}$ decreased, the intensities of the S-plasma peaks also decreased. While these peaks were faintly observed at $T_{Sulfur}$ = 100 °C, they became spectroscopically undetectable at $T_{Sulfur}$ = 90 °C. This indicates that the amount of S-plasma supply can be controlled by adjusting $T_{Sulfur}$.

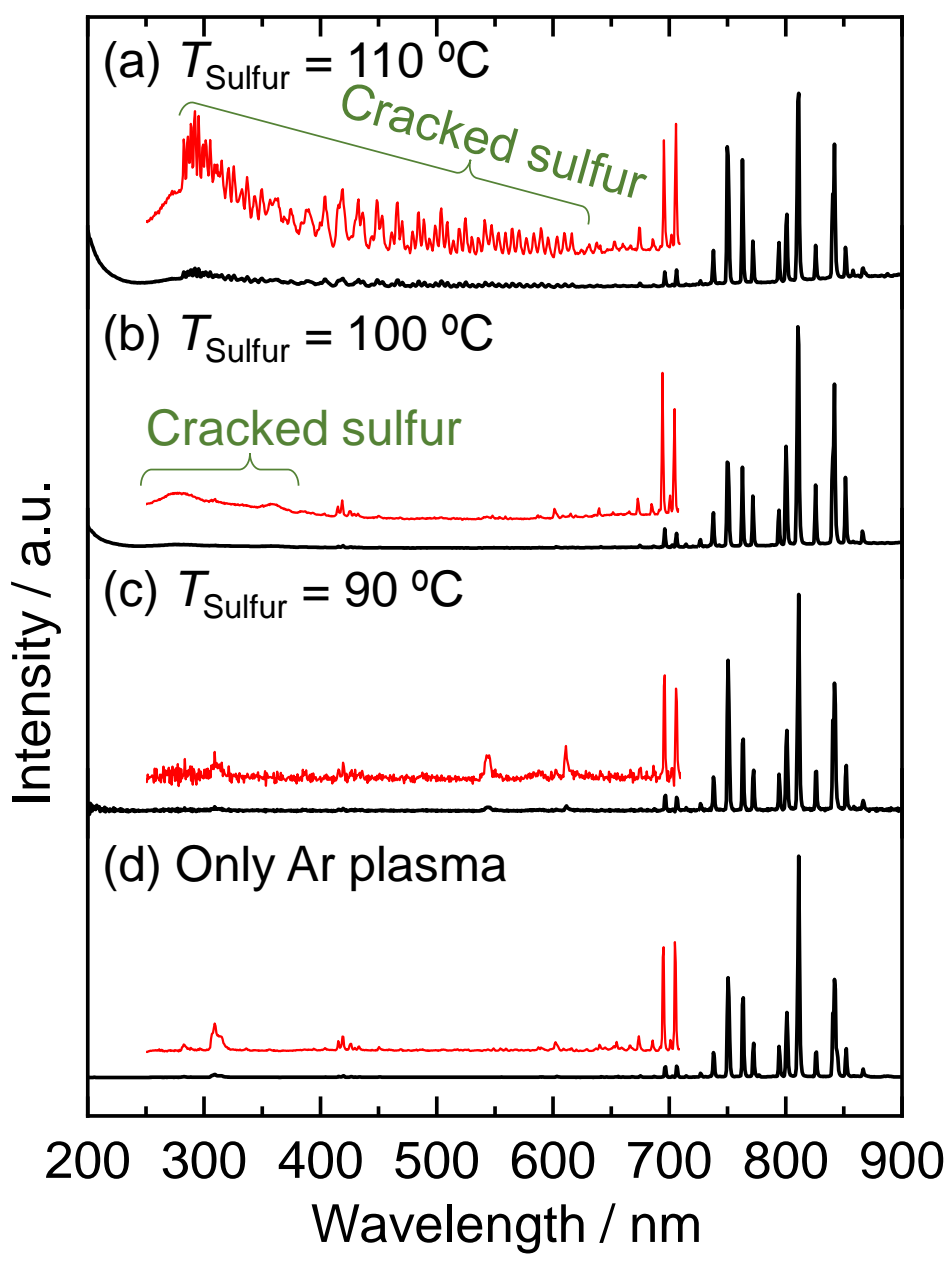

Figure S1. Emission spectra of the S-plasma (a) at $T_{Sulfur}$ = 110, (b) 100, and (c) 90 °C, and (d) the emission spectra of Ar plasma alone.

## Section S2. Analysis of Raman spectra

Raman spectra of the obtained SnS thin films exhibited four main peaks at approximately 95, 161, 190, and 217 $cm^{-1}$. The positions and FWHMs of these peaks (shown in Figure 3 of the main text) were determined by peak fitting using the Voigt function, as shown in Figure S2.

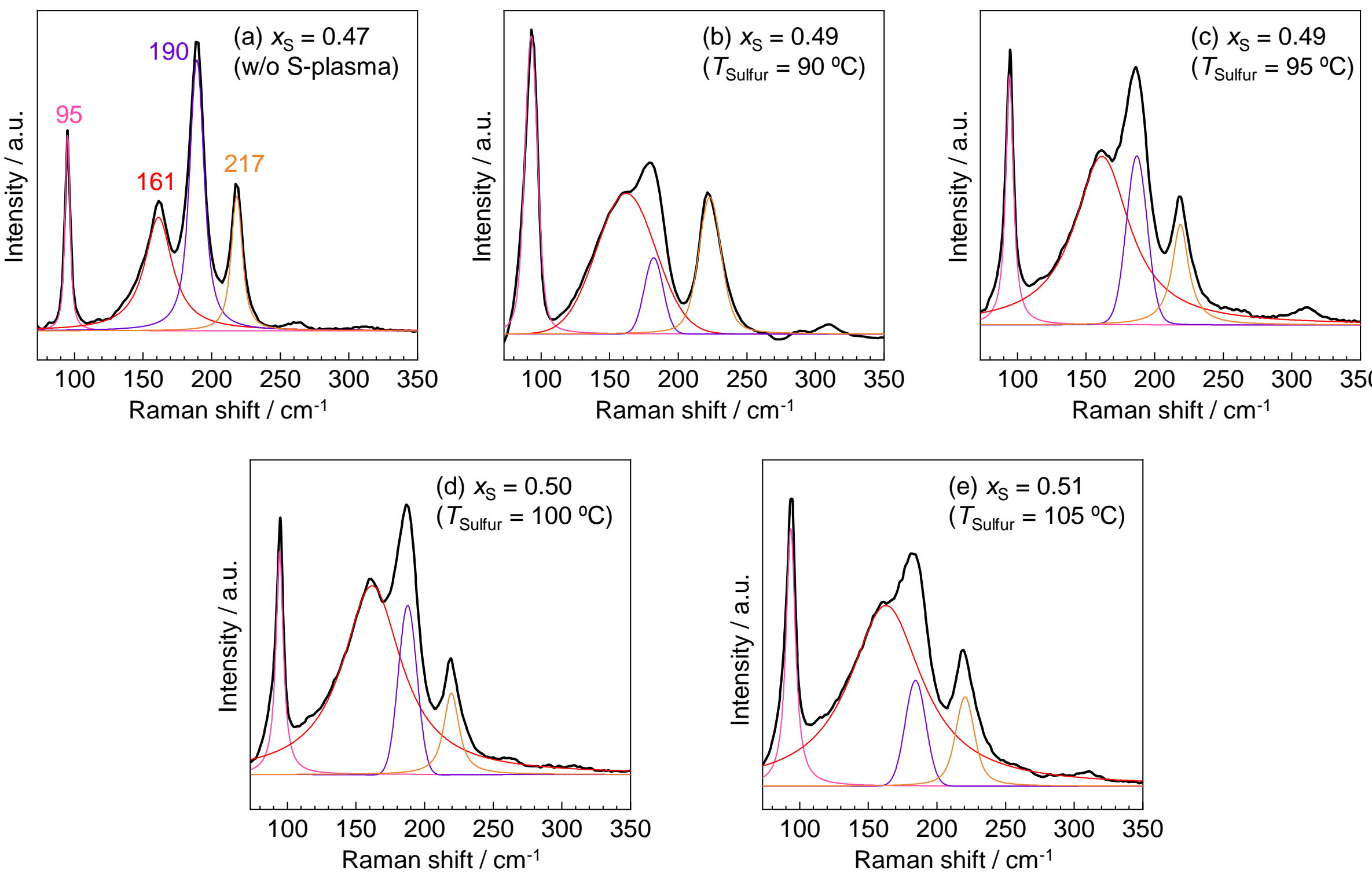

Figure S2. Peak fitting of Raman spectra of SnS thin films fabricated at (a) $T_{Sulfur}$ = 90, (b) 95, (c) 100, and (d) 105 °C. The spectra shown here were background-subtracted.

## Section S3. Detailed analysis of out-of-plane XRD of SnS thin films

Figure S3 shows the enlarged XRD profiles for the angle regions corresponding to the 200, 400, and 800 diffraction peaks of the SnS thin films fabricated at various $T_{\mathrm{Sulfur}}$. In the powder pattern of SnS, both the 111 and 400 diffraction peaks have high intensity ratios and similar diffraction angles, necessitating a careful distinction between them. A key point is that the 400 diffraction has a lower-order 200 diffraction, whereas the 111 diffraction does not. The presence of peaks at an angle corresponding to the 200 diffraction peak in all films fabricated with the S-plasma supply indicates that the peaks observed at 31–32° at least includes the 400 diffraction peak.

Thin films fabricated at $T_{\mathrm{Sulfur}}$ = 90 and 100 °C exhibited single peaks corresponding only to the 200, 400, and 800 diffractions. The blue line in the figure shows the peak positions determined under the assumption of a single lattice parameter; these positions match the observed peaks. Therefore, it can be concluded that only the 200, 400, and 800 diffraction peaks appeared in the out-of-plane XRD profiles for $T_{\mathrm{Sulfur}}$ = 90 and 100 °C.

In contrast, the thin films fabricated at $T_{\mathrm{Sulfur}}$ = 95 and 105 °C exhibited asymmetrical peaks at angles corresponding to the 400 and 800 diffractions. These XRD profiles were fitted using a split pseudo-Voigt function with XRD pattern analysis software (SmartLab Studio II v4.1, Rigaku Corp., Japan), as indicated by the blue and red peaks in the figures. The fitting results indicated that the peaks could be deconvoluted into two separate peaks. The peak observed on the higher-angle side was identified as the 111 diffraction peak because its lower-order diffraction did not appear at the corresponding angle of 16–17°, indicating that it was not a 400 diffraction peak. Conversely, the peak on the lower-angle side was identified as the 400 diffraction peak, because its lower-order 200 diffraction peak was observed as a single peak. From these results, it was concluded that the thin films fabricated at $T_{\mathrm{Sulfur}}$ = 90 and 100 °C predominantly contain crystallites oriented in the *h*00 direction, while the thin films fabricated at $T_{\mathrm{Sulfur}}$ = 95 and 105 °C contain crystallites oriented in both the *h*00 and 111 directions.

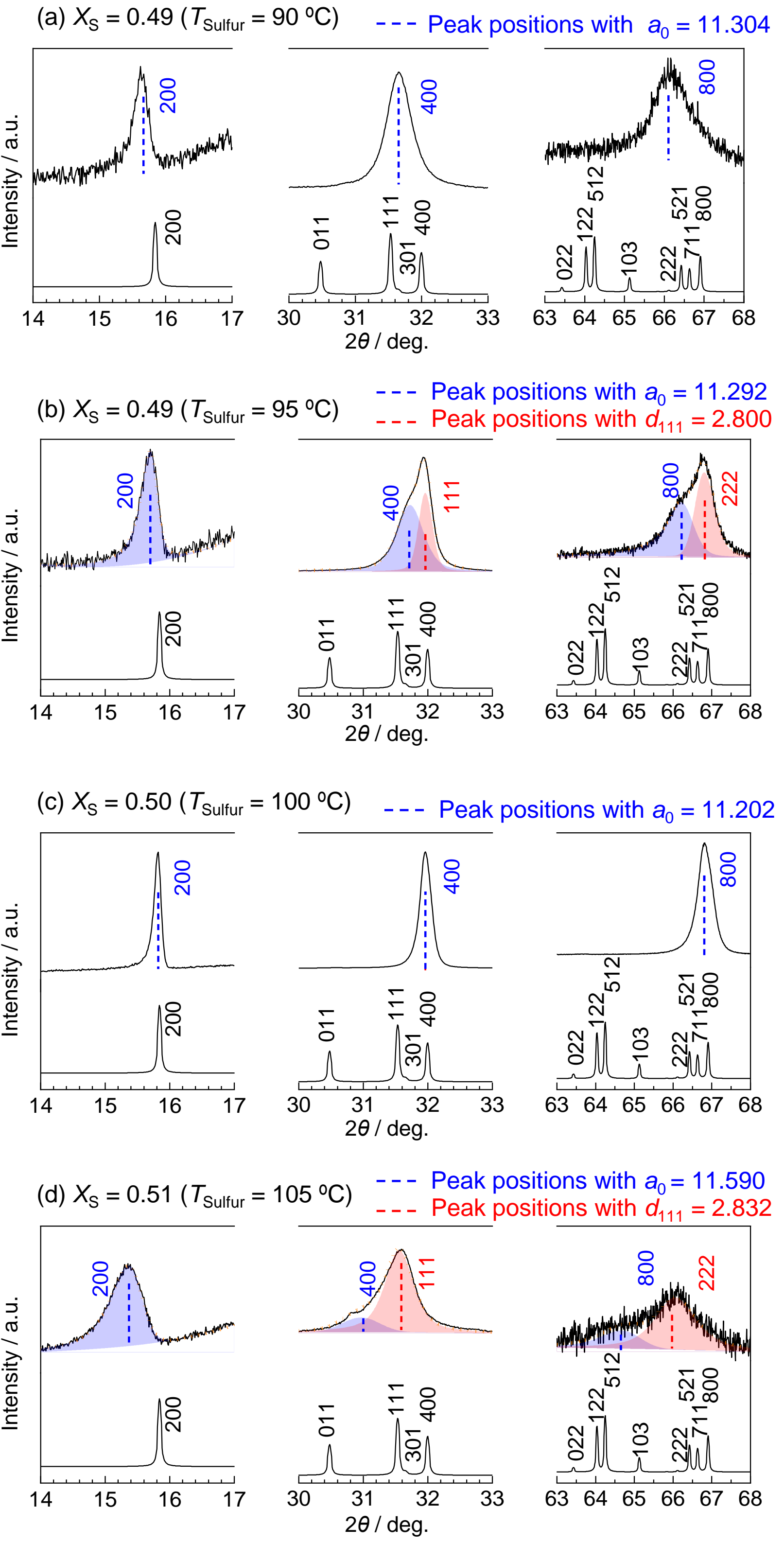

Figure S3. Enlarged XRD profiles of SnS thin films fabricated at (a) $T_{Sulfur}$ = 90, (b) 95, (c) 100, and (d) 105 °C. The blue or red vertical dotted lines indicate the corresponding peak positions for the assumed lattice parameter or lattice plane distance shown in the upper right corner of each figure.

## Section S4. Measurement of Seebeck coefficient

The Seebeck coefficients of the thin films were measured using a custom-made device, and the results are summarized in Table S1. One end of the thin film was in contact with a ceramic heater, while the other end was in contact with a Peltier element. The electromotive force was recorded using a multimeter. Typical measurement temperatures were 40 and 60 °C at each end. Au thin films were used as the electrodes.

Table S1. Compositional dependance of the Seebeck coefficient of thin films near room temperature.

| $T_{Sulfur}$ / °C | $x_S$ | Seebeck coefficient / μV $K^{-1}$ |
|---|---|---|
| w/o | 0.47 | +730 |
| 90 | 0.49 | +500 |
| 95 | 0.49 | +440 |
| 100 | 0.50 | +640 |
| 105 | 0.51 | +60 |

## Section S5. Temperature dependence of electrical conductivity

Figure S4 shows the temperature dependence of the electrical conductivity of the SnS thin films fabricated under various S-plasma supply conditions. All the thin films exhibited an almost linear relationship from −10 to −70 °C. The activation energies ($E_a$) were obtained from Arrhenius plots:

$$\sigma = \sigma_0 \exp\left(-\frac{E_a}{kT}\right)$$

where, $\sigma$, $\sigma_0$, $k$, and $T$ represent electrical conductivity, conductivity prefactor, Boltzmann constant, and absolute temperature, respectively.

Furthermore, the energy of the acceptor levels ($E_A - E_{VB}$, the energy difference between the valence band maximum and the acceptor level) was determined using the following equation:

$$E_A - E_{VB} = 2\,E_a$$

The results are summarized in Table S2.

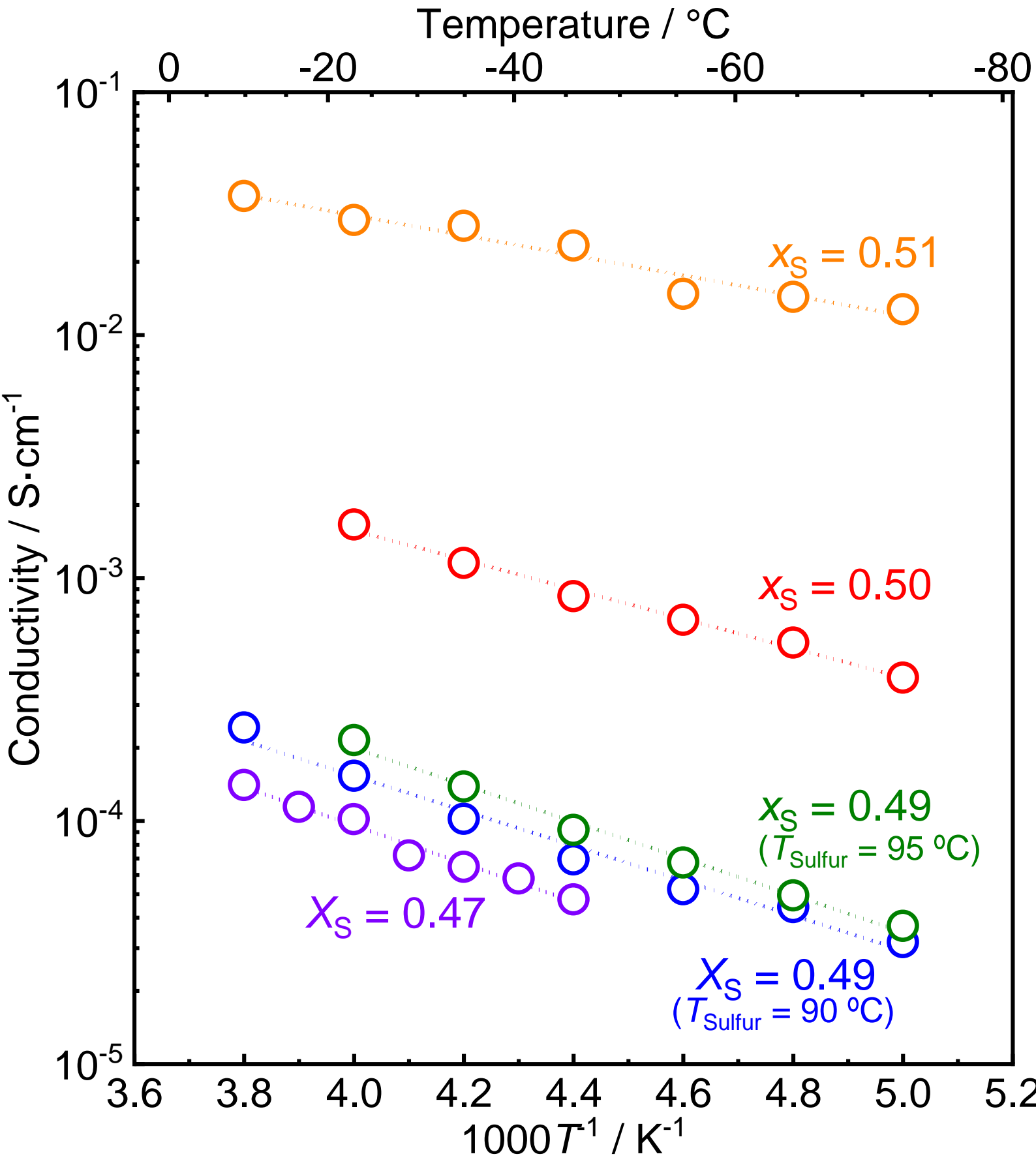

Figure S4. Temperature dependence of electrical conductivity of SnS thin films with various compositions.

Table S2. Activation energy ($E_a$) and energy of acceptor level ($E_A - E_{VB}$).

| $T_{Sulfur}$ / °C | $X_S$ | $E_a$ / meV | Acceptor level, $E_A - E_{VB}$ / meV |
|---|---|---|---|
| w/o | 0.47 | 156 | 312 |
| 90 | 0.49 | 166 | 332 |
| 95 | 0.49 | 161 | 322 |
| 100 | 0.50 | 107 | 214 |
| 105 | 0.51 | 82 | 164 |

## Section S6. Analysis of spectroscopic ellipsometry spectra

The delta and phi spectra of the stoichiometric SnS thin film fabricated at $T_{Sulfur}$ = 100 °C were measured using a spectroscopic ellipsometer, and an optical model that could reproduce these spectra was constructed. As a good fit was not achieved with the uniform single-layer model, a surface layer was added as the top layer of the SnS thin film to account for surface contamination, oxidation, and morphological roughness. As shown in Figure S5, the model which includes a 23 nm surface layer on top of a 198 nm bulk SnS layer provided a good fit with a determination coefficient ($R^2$) of 0.99644. The absorption coefficient spectrum shown in Figure 12 of the main text represents the absorption coefficient of the bulk SnS layer without the surface layer. Good fits were not achieved for the non-stoichiometric thin films (i.e., $T_{Sulfur}$ = 90, 95, and 105 °C), likely due to the sparse morphology (see Figure 8 of the main text), which significantly influenced their reflection.

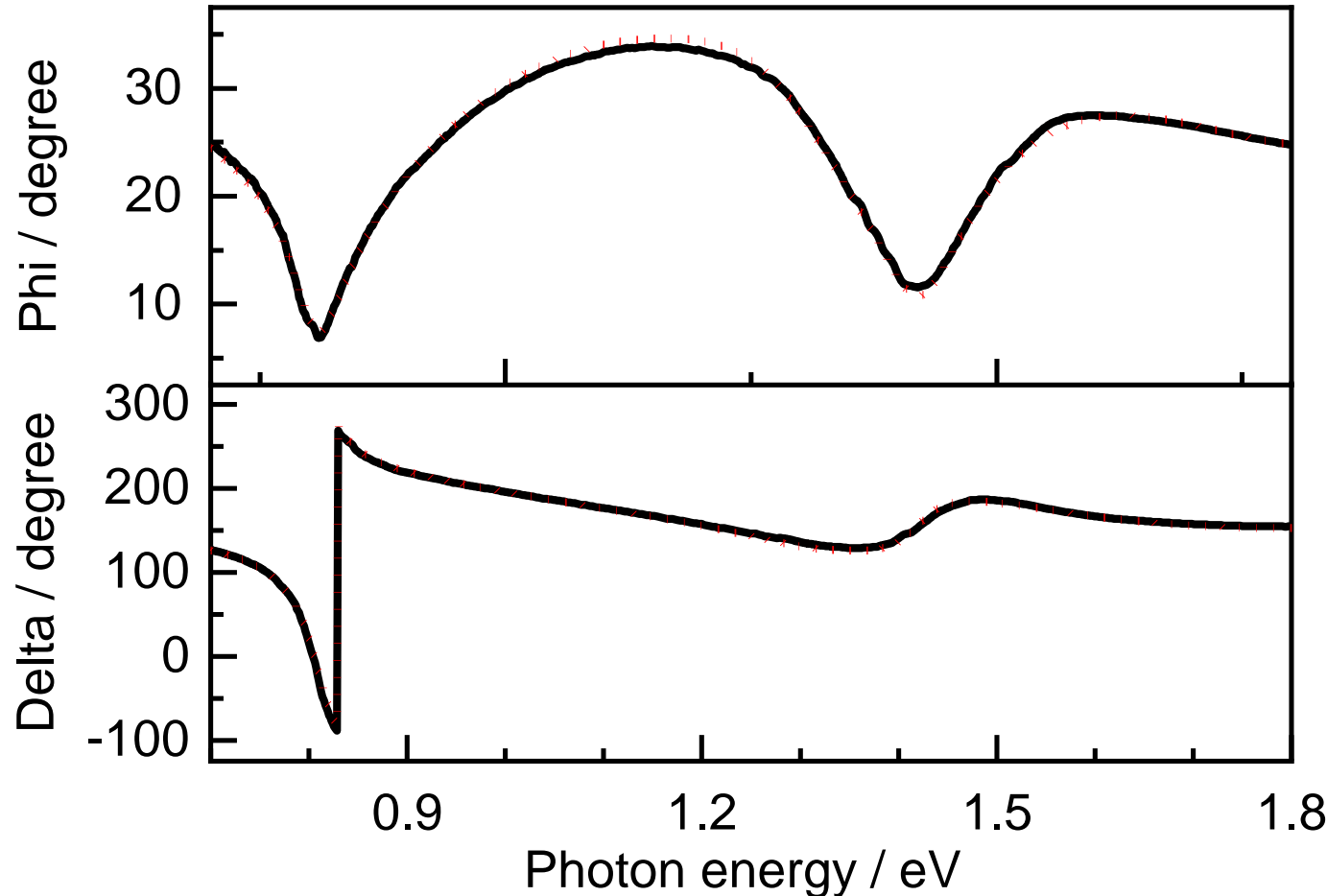

Figure S5. Phi and Delta spectra from ellipsometry measurements (black solid line) and fitting results (red dotted line).

## References for supporting information